\documentclass[11pt]{article}
\pdfoutput=1
\usepackage[T1]{fontenc}
\usepackage{lmodern}
\usepackage{smile}
\usepackage{fullpage}
\usepackage[protrusion=true,expansion=true]{microtype}
\usepackage[dvipsnames]{xcolor}
\usepackage[colorlinks,linkcolor=BrickRed,citecolor=NavyBlue,urlcolor=NavyBlue]{hyperref}
\usepackage{tabularx}
\usepackage{enumitem}
\usepackage{setspace}
\usepackage{caption}
\usepackage{tikz}
\usetikzlibrary{positioning,arrows.meta}
\usepackage{placeins}
\setlist{nosep,leftmargin=*}
\graphicspath{{figures/}}
\setcitestyle{authoryear,round}

\newcommand{\pool}{\mathcal{P}}
\newcommand{\lineups}{\mathcal{L}}
\newcommand{\draws}{\mathcal{D}}
\newcommand{\cells}{\mathcal{C}}
\newcommand{\Rt}{\mathcal{R}}
\newcommand{\Own}{\mathcal{O}}
\newcommand{\FA}{\mathcal{F}}
\newcommand{\Tr}{\mathcal{T}}
\newcommand{\pts}{\text{pts}}

\newcommand{\VSS}{\mathrm{VSS}}

\title{Shooting for Fit: Risk-Averse Stochastic Roster Construction from Bayesian Shot-Pattern Models}
\author{
  Yue Yu\thanks{Department of Statistics, Indiana University; email: {\tt yyu3@iu.edu}}
  \and
  Cong Zhang\thanks{School of Economics, University of Nottingham Ningbo; email: {\tt cong.zhang@nottingham.edu.cn}}
  \and
  David Crandall\thanks{Department of Computer Science, Indiana University; email: {\tt djcran@iu.edu}}
  \and
  Guanyu Hu\thanks{Department of Statistics and Probability, Michigan State University; email: {\tt huguanyu@msu.edu}}
}
\date{}

\begin{document}
\maketitle

\begin{abstract}
Roster construction is a critical decision problem in professional basketball, requiring teams to balance player quality, player complementarity, and financial constraints under uncertainty. Existing roster optimization models primarily rely on individual player ratings and aggregate performance statistics, which provide limited information about how players' offensive contributions vary with their teammates. Recent advances in player-tracking technology have enabled statistical analyses of spatial shooting patterns, but these insights have rarely informed roster construction. We bridge this gap by developing a data-driven, risk-averse two-stage stochastic optimization framework for NBA roster construction. We first develop a Bayesian modeling framework that combines player statistics and spatial shot-chart data to identify player roles and estimate how teammate composition influences shot selection, shooting frequency, and scoring efficiency. The resulting model provides uncertainty-quantified offensive values for hypothetical five-player lineups, including combinations that have never played together. We then formulate roster construction as a two-stage stochastic mixed-integer program that optimizes player acquisitions under NBA salary-cap regulations while allowing lineup playing-time allocations to adapt to uncertain player roles. We evaluate the framework using 240 team-offseason decisions over eight seasons and two additional held-out seasons. The results demonstrate that incorporating spatial shooting compatibility substantially changes optimal roster decisions and improves predicted offensive performance relative to approaches based on individual player values. We further find that flexibility in allocating playing time across lineups absorbs much of the uncertainty about player roles. Our findings highlight how detailed spatial shooting information can be translated into actionable roster decisions through stochastic optimization.
\end{abstract}


\section{Introduction}\label{sec:intro}

Constructing a competitive roster is one of the most consequential decisions in professional basketball. Each offseason, National Basketball Association (NBA) teams must determine which players to retain, acquire, or release while balancing individual player quality, team compatibility, and complex financial constraints. These decisions have motivated a substantial literature in operations research, where roster construction is typically formulated as a constrained optimization problem involving player selection, performance evaluation, and resource allocation \citep{boon2003,pantuso2017,muniz2023}. However, the success of a roster depends not only on the individual abilities of its players but also on how players' offensive contributions reinforce or constrain one another. In July 2019, for example, the Los Angeles Clippers signed Kawhi Leonard, the most valuable player of the previous NBA Finals, and traded a promising young guard, a starting forward, and five first-round draft picks to acquire Paul George, who had finished third in MVP voting the previous season. Although the pairing was widely expected to produce a championship contender, it ultimately failed to deliver a title. This example illustrates a fundamental challenge in roster construction: how to identify player combinations in which individual abilities reinforce one another and translate into greater value for the team as a whole.

A central difficulty is that basketball performance is inherently interactive. With only one ball available, the offensive opportunities of one player depend on the tendencies and decisions of his teammates. In this sense, players can act as substitutes or complements in offensive production. Two players who favor similar shooting locations may compete for scoring opportunities, whereas a perimeter shooter may create additional space for teammates who attack the basket. A playmaker who attracts defensive attention may also improve the quality of shots available to others. Consequently, a player's marginal contribution to team performance depends on the composition of the roster, and the same player may contribute substantially different value across different rosters. Previous studies have recognized the importance of player interactions through lineup synergy, adjusted player ratings, and passing networks \citep{maymin2013,bu2019,kalman2020,grassetti2021,yamada2024}. Nevertheless, anticipating how players will perform together remains difficult, particularly when evaluating potential acquisitions who have never played with the current roster. These decisions must also be made before the season begins and under salary-cap regulations that restrict feasible player combinations. In a typical season, players new to a team account for approximately one-third of its playing minutes, while a single star contract can consume a substantial share of the team's payroll. Effective roster construction therefore requires a framework that accounts for player compatibility, financial feasibility, and uncertainty about future lineup performance.

A substantial body of operations research has developed optimization approaches to roster construction and related player selection problems. Early models primarily relied on traditional player information, including positions, individual ratings, and aggregate performance statistics. For example, \citet{boon2003} match the quality required at each position with the quality supplied by available players. In football, \citet{pantuso2017} formulate squad composition as a stochastic optimization problem with uncertain player values, while \citet{pantuso2021} incorporate chance constraints to maintain financial sustainability. More recently, \citet{beyondpitch2026} combine machine learning and optimization to construct financially sustainable squads. In baseball, \citet{chan2019} investigate the value of players who can cover multiple positions, while \citet{haugh2021} formulate daily-fantasy lineup selection as a portfolio optimization problem under uncertainty. Basketball-specific studies include \citet{muniz2023}, who optimize draft selections, trades, and free-agent acquisitions while accounting for synergy between player archetypes, and \citet{zhang2026}, who develop a multiseason franchise management framework using distributionally robust optimization. Related research has examined individual player valuation \citep{muller2017,mchale2018,kharrat2020}. More broadly, roster construction shares important features with workforce planning under heterogeneous skills \citep{debruecker2015}, where long-term personnel commitments must be coordinated with flexible operational assignments \citep{malaki2023}. Despite these developments, club-level roster decisions have received comparatively less attention than league-level optimization problems involving scheduling, tournament design, seeding, and ranking \citep{devriesere2025}.

Although existing optimization models provide useful tools for player selection, their evaluation of candidate rosters is generally based on individual player values or relatively coarse measures of compatibility. Player contributions are often represented by aggregate statistics, such as scoring averages, shooting percentages, and player ratings, while interactions are captured through predefined player types or pairwise synergy parameters. Such representations may overlook important differences in offensive playing styles. For example, two players with similar scoring averages and shooting percentages may have very different spatial shooting patterns. One may primarily attempt shots near the basket, whereas another may rely heavily on perimeter jump shots. Differences in spatial shooting patterns can substantially affect player compatibility, even when players have similar aggregate statistics. Moreover, teammate composition may influence not only where a player shoots but also how frequently he shoots and how efficiently he scores. Understanding these effects is essential for evaluating hypothetical lineups, yet conventional player-level summary statistics provide limited information about them.

Recent developments in player-tracking technology and the availability of high-resolution basketball data offer new opportunities to address these limitations. Modern tracking systems and detailed play-by-play records provide increasingly rich information about shot locations, shooting outcomes, player movements, and the identities of teammates and opponents on the court. In particular, spatial shot-chart data characterize where players attempt shots and how successfully they score across different court locations, providing a more detailed representation of offensive behavior than traditional aggregate statistics. These advances have stimulated a growing statistical literature on the spatial analysis of basketball performance. For example, \citet{reich2006}, \citet{miller2014}, and \citet{jiao2021} develop statistical models to characterize spatial shooting patterns, while \citet{hu2021}, \citet{wongtoi2023}, and \citet{yin2023} investigate clustering methods for identifying players with similar shot-selection behavior. Other studies develop spatial models that borrow information across neighboring court locations \citep{yin2022} or evaluate player contributions through expected points \citep{williams2025}.

The availability of player-tracking data has also enabled more detailed investigations of offensive and defensive interactions. For instance, \citet{franks2015} and \citet{cervone2016} use spatial tracking information to study defensive effects and possession values, while \citet{sandholtz2020} examine the offensive consequences of suboptimal shot allocation. These developments demonstrate that high-resolution basketball data can reveal aspects of individual performance and team interactions that are not apparent from conventional summary statistics. From a modeling perspective, the composition of teammate roles can also be characterized using compositional data, for which the effects of individual components are interpreted relative to the others \citep{li2023,meng2025,deng2026linking}. Recent developments in compositional regression further allow components with similar effects to be identified and grouped \citep{xu2026}.

Despite substantial progress in statistical modeling of basketball shot charts and player-tracking data, these developments have not been fully incorporated into roster construction decisions. Existing shot-chart studies primarily focus on modeling spatial shooting behavior, clustering players, evaluating individual performance, or understanding lineup interactions. In contrast, the roster construction literature continues to rely largely on aggregate player characteristics and summary performance measures. In particular, while previous approaches have considered compatibility through player archetypes or observed lineup outcomes, they do not directly exploit spatial shooting patterns to predict how a player's shot selection, shot frequency, and shooting efficiency vary with teammate combinations. This creates a gap between the information available from modern basketball tracking systems and the information used in roster optimization. Bridging this gap requires a framework that translates spatial shooting behavior into the offensive value of hypothetical lineups and integrates those values into constrained roster decisions.

In this paper, we address this gap by developing an integrated statistical and optimization framework that uses spatial shot-chart data to guide NBA roster construction. To the best of our knowledge, this is the first study to directly incorporate spatial shot-chart information into a roster optimization framework for evaluating hypothetical player combinations and making acquisition decisions. Our central idea is to estimate how each player's offensive contribution varies with teammate composition, rather than treating individual player values as fixed across potential rosters. By linking spatial statistical modeling with stochastic optimization, the proposed framework translates detailed player-level shooting information into actionable roster recommendations.

\begin{figure}[t]
\centering
\definecolor{dataC}{HTML}{8C6D1F}\definecolor{dataF}{HTML}{F7F1E3}
\definecolor{statC}{HTML}{1F3A5F}\definecolor{statF}{HTML}{E6EDF5}
\definecolor{decC}{HTML}{882255}\definecolor{decF}{HTML}{F5E6EE}
\definecolor{evalC}{HTML}{2A8C82}\definecolor{evalF}{HTML}{E2F1EF}
\begin{tikzpicture}[font=\footnotesize,
  box/.style={draw=#1, line width=0.7pt, fill=white, rounded corners=2pt, align=center, text width=3.05cm,
              minimum height=1.32cm, inner sep=2.5pt, execute at begin node={\hyphenpenalty=10000\exhyphenpenalty=10000}},
  lanetitle/.style={align=center, text=#1, font=\footnotesize\bfseries},
  arr/.style={-{Stealth[length=1.9mm]}, line width=0.7pt, black!55},
  lab/.style={font=\scriptsize\itshape, text=black!65, fill=white, inner sep=1pt}]
\foreach \x/\f in {0/dataF, 4.2/statF, 8.4/decF, 12.6/evalF}
  \fill[\f, rounded corners=4pt] (\x,0.62) rectangle ++(3.75,-6.12);
\node[lanetitle=dataC]  at (1.875,0.22)  {Data, season $t$\\[-1pt]{\normalfont\scriptsize Section~\ref*{sec:data}}};
\node[lanetitle=statC]  at (6.075,0.22)  {Statistical model\\[-1pt]{\normalfont\scriptsize
  Section~\ref*{sec:model-value}}};
\node[lanetitle=decC]   at (10.275,0.22) {Decision model\\[-1pt]{\normalfont\scriptsize
  Sections~\ref*{sec:model-program} and~\ref*{sec:model-solve}}};
\node[lanetitle=evalC]  at (14.475,0.22) {Evaluation\\[-1pt]{\normalfont\scriptsize Sections~\ref*{sec:results}
  and~\ref*{sec:cases}}};
\node[box=dataC] (box)    at (1.875,-1.25) {Box scores, minutes\\and positions\\of every player};
\node[box=dataC] (shots)  at (1.875,-3.0)  {Every shot: where,\\how it was created,\\made or missed};
\node[box=dataC] (stints) at (1.875,-4.75) {Every stint: the ten\\players on the floor};
\node[box=statC] (roles)  at (6.075,-1.25) {\textbf{Roles}: seven,\\learned by Bayesian\\clustering};
\node[box=statC] (shotm)  at (6.075,-3.0)  {\textbf{Shot model}:\\teammates shift\\attempts and makes};
\node[box=statC] (values) at (6.075,-4.75) {\textbf{Lineup values} $V_{\ell d}$\\in each role draw $d$};
\node[box=decC] (rules)   at (10.275,-1.25) {Contracts and the\\salary rules of\\season $t+1$};
\node[box=decC] (stage1)  at (10.275,-3.0)  {\textbf{Stage 1}, summer:\\keep, re-sign,\\sign, trade};
\node[box=decC] (stage2)  at (10.275,-4.75) {\textbf{Stage 2}, season\\$t+1$: minutes in\\each role draw};
\node[box=evalC] (real)   at (14.475,-1.25) {Plans against the\\real rosters of\\240 offseasons};
\node[box=evalC] (simple) at (14.475,-3.0)  {Plans against those\\of simpler models};
\node[box=evalC] (held)   at (14.475,-4.75) {Two held-out seasons\\and three famous\\summers};
\draw[arr] (box) -- (roles);
\draw[arr] (shots.north east) -- (roles.south west);
\draw[arr] (shots) -- (shotm);
\draw[arr] (stints.north east) -- (shotm.south west);
\draw[arr] (roles) -- (shotm);
\draw[arr] (shotm) -- (values);
\draw[arr] (rules) -- (stage1);
\draw[arr] (stage1) -- node[lab, right=2pt] {roster} (stage2);
\draw[arr] (values) -- node[lab, above=2pt] {values} (stage2);
\draw[arr] (stage2.east) -- (held.west);
\draw[arr] (stage2.east) -- ++(0.27,0) |- (simple.west);
\draw[arr] (stage2.east) -- ++(0.27,0) |- (real.west);
\end{tikzpicture}
\caption{The engine for one team-offseason, in four layers. Data from season $t$ feed the statistical models, whose
lineup values enter the two-stage stochastic program; its plans are then compared with the rosters teams built and with
the plans of simpler models. The models are fitted on data up to season $t$, and the posterior draws of the roles are the
scenarios of the program.}\label{fig:pipeline}
\end{figure}
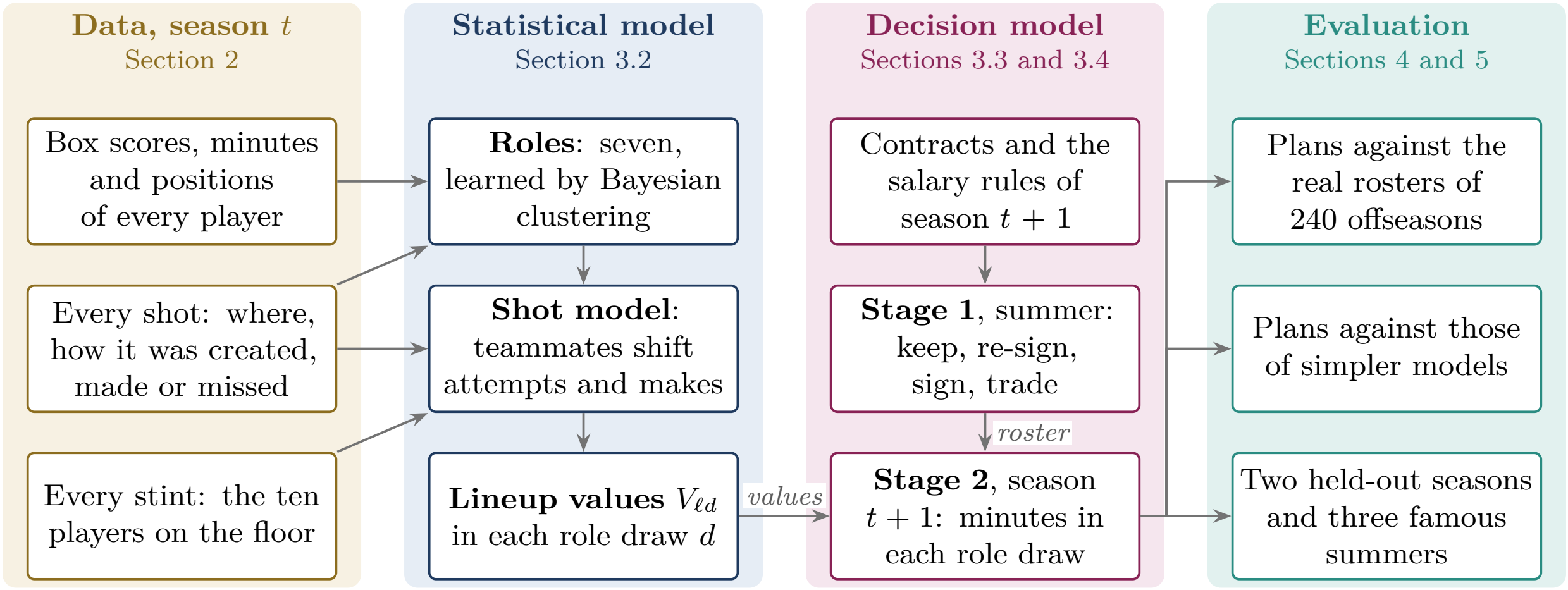

The proposed approach follows the decision-making process of an NBA front office, as illustrated in Figure~\ref{fig:pipeline}. We first characterize players using offensive roles learned from traditional performance statistics and spatial shot charts from the previous season. These roles provide transferable representations of playing styles and enable the evaluation of potential acquisitions who have never played with the current roster. We then develop a Bayesian shot-pattern model that estimates how the roles of a player's four teammates influence his shot locations, shot frequency, and shooting efficiency. The estimated relationships are used to construct an offensive value function for any hypothetical five-player lineup, expressed in expected points per 100 possessions, while also accounting for posterior uncertainty about its performance.

We subsequently incorporate these lineup values into a risk-averse two-stage stochastic mixed-integer optimization model. In the first stage, the model determines which players to retain or acquire subject to salary-cap regulations and transaction constraints. In the second stage, it allocates playing time across feasible lineups under different realizations of uncertain player roles. This formulation distinguishes long-term roster commitments from the operational flexibility available to coaches, who can adjust lineup rotations after observing how players perform together. Consequently, the framework jointly accounts for spatial shooting compatibility, uncertainty in player roles, and the flexibility of playing-time allocation.

We evaluate the proposed framework through a retrospective analysis of roster decisions for all 30 NBA teams across eight offseasons from 2016 to 2023, yielding 240 team-offseason decision problems. For each offseason, the model uses only information available at the time of the decision and compares its recommended roster with the roster actually assembled through trades and free-agent signings. To assess out-of-sample generalizability, we further evaluate the framework on two additional seasons held out throughout model development, using a frozen implementation. The empirical findings highlight the importance of accounting for teammate compatibility when constructing competitive rosters. Compared with methods that primarily value players using individual performance measures, our framework recommends substantially different rosters with higher predicted offensive performance. The estimated improvements would move a league-average offense to about sixth in the league. We also find that flexibility in allocating playing time across lineups allows teams to accommodate much of the uncertainty associated with player roles. A case study of the 2019 Los Angeles Clippers further illustrates how spatial shooting compatibility can change the evaluation of high-profile roster decisions.

Our study makes three main contributions. First, we bridge the gap between statistical analysis of spatial shot charts and roster optimization by developing a shot-chart-driven lineup valuation framework. Unlike existing roster construction approaches that primarily rely on aggregate player ratings or coarse synergy measures, our model explicitly characterizes how teammate composition affects shot selection, shot frequency, and shooting efficiency. This provides a transferable framework for evaluating hypothetical lineups, including combinations of players who have never played together, while quantifying uncertainty in their offensive performance.

Second, we develop a risk-averse two-stage stochastic mixed-integer optimization framework that converts shot-based lineup values into roster construction decisions. The first stage optimizes player acquisitions under the salary regulations of three collective bargaining agreements, while the second stage adjusts playing-time allocations across uncertain scenarios. This formulation integrates player compatibility, financial constraints, and adaptive lineup management within a unified decision framework.

Third, we provide a comprehensive empirical evaluation using 240 historical team-offseasons and two additional held-out seasons. Through comparisons with actual front-office decisions, predictive benchmarks, and component-level analyses, we quantify the decision value of spatial shooting information, teammate compatibility, uncertainty modeling, and lineup flexibility. These results illustrate how high-resolution sports data can improve not only statistical evaluation but also downstream optimization decisions.

From a methodological perspective, our framework builds on the predict-then-optimize paradigm \citep{bertsimas2020,elmachtoub2022} by integrating Bayesian predictions with stochastic optimization. Rather than training the predictor directly for downstream optimization, we investigate which components of a rich predictive model materially change the resulting decisions. We quantify their decision value using established measures from stochastic programming \citep{birge2011}. Posterior draws of player roles generate scenarios for sample average approximation \citep{kleywegt2002}, while a tail-risk objective protects against unfavorable outcomes \citep{rockafellar2000}. We also examine the robustness of recommended rosters to defensive adjustments through a budgeted uncertainty set \citep{bertsimas2004}.

The remainder of the paper is organized as follows. Section~\ref{sec:data} describes the data and research questions. Section~\ref{sec:model} presents the shot-pattern model, lineup valuation, and stochastic optimization framework. Section~\ref{sec:results} evaluates the proposed approach using historical and held-out NBA seasons. Section~\ref{sec:cases} examines three case studies involving major roster decisions. Section~\ref{sec:conclusion} discusses managerial implications and concludes the paper. Appendices~\ref{app:data}--\ref{app:protocol} detail the data, the role and shot models, the salary-rule constants, the calibration of the stress test and the frozen evaluation protocol.

\section{Data and research questions}\label{sec:data}

\subsection{Data}\label{sec:data-sources}

All data used in this study are public. Play-by-play logs, shot locations, box scores and rotation records come from
NBA.com/Stats, the league's official statistics website \citep{nbastats}, accessed through the Python client nba\_api
\citep{patel2018}, for eleven regular seasons, 2015-16 to 2025-26, and the playoffs of 2015-16 to 2024-25; the nine
seasons from 2015-16 to 2023-24 alone hold 1.88 million shots. On-court lineups are reconstructed from rotation and
play-by-play records, so that every shot is linked to the ten players on the floor. Salaries, players' moves and
contract status, team offensive ratings and playoff results come from Basketball-Reference \citep{bbref}; the salaries
of 2024-25 and 2025-26 come from its contracts page as archived after each season's opening night
\citep{bbrefcontracts}. The amounts of the salary cap, the tax line and the exceptions of each league year come from the
league's press releases \citep{nbapr} and contemporary press reports \citep{hoopsrumors,bbinsiders}, and the salary rules
from the collective bargaining agreements of 2011, 2017 and 2023 \citep{nbacba2011,nbacba2017,nbacba2023}.

To evaluate how teammates affect a player's opportunities, we describe every shot by where it is taken and how it is
created. The court is divided into five zones (the restricted area at the rim, the rest of the paint, mid-range, the
corner three and the above-the-break three). Each shot is categorized by its movement type: a drive to the basket, a
cut, a pull-up jumper off the dribble, a post-up or touch in the paint, or a spot-up shot off a pass. Crossing the five
zones with the five movement types yields 25 \emph{cells}. Movement type separates the shots a player creates for himself
from the shots created by his teammates: a corner three off a pass and a pull-up three from that spot depend on very
different teammates. Zones are calculated from recorded shot coordinates, and movement types come from play-by-play
descriptions, which we checked season by season against the league's player-tracking data (\citealp{nbastats}; Appendix~\ref{app:data}).

\begin{figure}[tbp]
\centering
\subfigure[Where the league's shots come from, for all attempts and for each movement type, in 2022-23.
Darker hexagons hold a larger share of the panel's attempts.]{\includegraphics[width=0.85\textwidth]{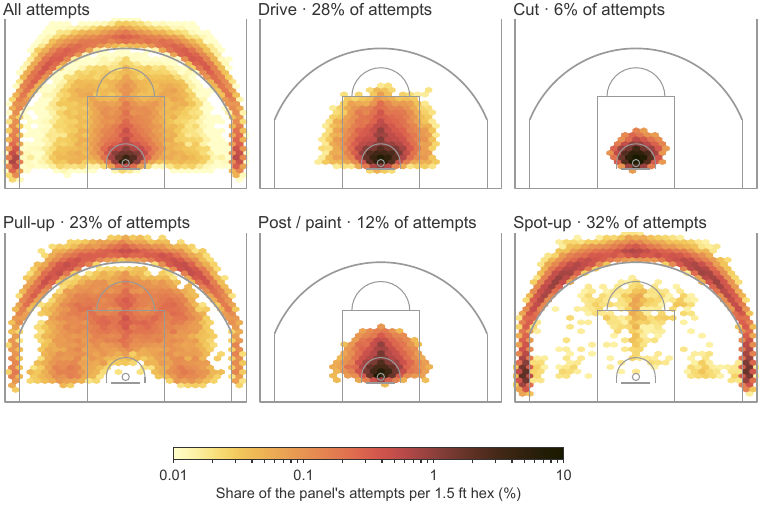}
\label{fig:data-league}}\\
\subfigure[One representative player of each role of Table~\ref{tab:roles} in 2022-23: every shot, by movement type, made (filled) or
missed (open). The bar under each court gives the player's mix of movement types.]{\includegraphics[width=0.85\textwidth]{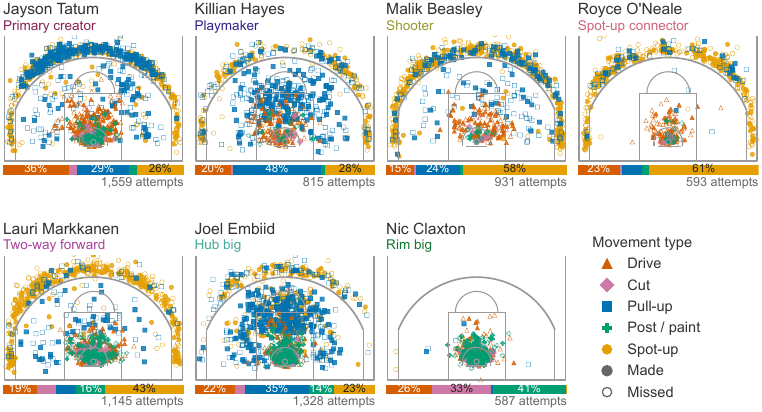}
\label{fig:data-players}}
\caption{The shot data.}\label{fig:data}
\end{figure}

The shot data show why offensive production cannot be evaluated by adding individual shot charts together. Each
movement type has its own geography (Figure~\ref{fig:data-league}): drives and cuts end near the basket, spot-ups fill
the corners and the arc, and pull-ups spread across the half court. Individual players differ even more
(Figure~\ref{fig:data-players}). Some take most of their shots off the dribble across all zones, others wait for spot-ups
in the corners, and others finish at the rim. Because five players share one ball, these individual tendencies cannot
simply be added up: how many shots a player takes in any cell depends directly on whether his teammates want those same
shots.

The decision engine is evaluated over eight pairs of seasons. Models estimated on one season make the decisions for the
following summer, from 2015-16 $\to$ 2016-17 to 2022-23 $\to$ 2023-24. Two later seasons, 2024-25 and 2025-26,
were kept aside throughout model development: nothing from them was opened until the entire method had been frozen,
providing a clean final test (Section~\ref{sec:results-sealed}).

\subsection{Research questions}\label{sec:questions}

The patterns in the shot data motivate the four research questions this paper addresses. Because players differ
sharply in where and how they shoot, and because five players share one ball, a newcomer inevitably changes the shots of
the teammates already on the roster. A front office must therefore determine whether these mutual adjustments can be
predicted for lineups that have never played together, whether acting on those predictions leads to better rosters, and
which components of the model actually alter the resulting decisions. We investigate four questions in sequence:

\begin{enumerate}[label=\textbf{Q\arabic*.},leftmargin=2.6em]
\item \emph{Can teammate fit be predicted?} Do the roles of a player's teammates shift his shots in ways that hold in
the next season, carry over to players who change teams, and track how well teams score?
(Section~\ref{sec:results-forecast})
\item \emph{Does planning for fit build better rosters?} Given the same information, options and salary rules, does the
engine's plan improve on the roster each team actually assembled, and does the improvement hold in seasons kept aside?
(Sections~\ref{sec:results-rosters} and~\ref{sec:results-sealed})
\item \emph{Which parts of the model change the decision?} How much does a front office lose if it ignores teammate
effects, plans for each player's average role instead of the uncertainty about roles, ignores how teammates crowd the
same shots, or ignores playoff defenses, and how many rosters are nearly as good as the best?
(Section~\ref{sec:results-drivers})
\item \emph{What does the engine say about decisions that shaped the league?} How does it read Cleveland's summer of
2017, the Clippers' of 2019, Brooklyn's of 2020 and the moves of 2023, and what lies outside its view?
(Section~\ref{sec:cases})
\end{enumerate}

The first question evaluates the statistical shot model, while the remaining three address the optimization decisions.
Throughout, a richer model earns its place only if the plans it produces differ from, and score better than, the plans of
a simpler one.

\section{The roster construction model}\label{sec:model}

This section builds the decision model for one team-offseason. Section~\ref{sec:model-problem} describes the decision a
front office faces, Section~\ref{sec:model-value} turns shot patterns into the value of any lineup,
Section~\ref{sec:model-program} formulates the two-stage stochastic program, and Section~\ref{sec:model-solve} explains
how we solve the program and evaluate its plans.

\subsection{Problem description}\label{sec:model-problem}

An NBA offseason forces a front office to make binding personnel commitments before the coaching staff can observe the
new players on the court. Every summer, once the playoffs end, each team enters a transaction window. The front office
holds the contracts of some players, decides whether to re-sign its own free agents, and pursues outside signings or
trades. Three practical questions govern this summer planning: which incumbent free agents to retain, whether to spend
cap space or over-the-cap exceptions, and which trade packages satisfy the league's salary-matching rules. Once the
season opens, the head coach distributes playing time across five-player lineups as he observes which player
combinations work well together.

This operational calendar creates a natural two-stage decision structure separated by time and information. During the
summer offseason, the front office sets the roster for season $t+1$ without knowing how incoming and returning players
will blend on the court. During the subsequent season, the coach observes on-court chemistry and adjusts playing time
across lineups. Our mathematical formulation follows the same sequence. We estimate all components from data up to season $t$, including player roles, the shot model, and candidate lineup values. The first stage selects the roster for
season $t+1$ under the collective bargaining rules of the coming league year. The second stage determines the playing
time of each lineup across $D=20$ posterior draws of player roles, where each draw represents a plausible scenario of
how the players will be used. In stochastic programming terms, the roster represents the first-stage
decision, lineup minutes provide the recourse, and the posterior role draws serve as scenarios. We solve each
team-offseason as a single program across all scenarios simultaneously. Because these scenarios originate directly from
the Bayesian role model, they express the model's own uncertainty about how players will be used; they are not noise
added for the optimization.

To define the available personnel choices, we assemble the team's candidate options after season $t$ into a pool
$\pool$. This pool includes all players under contract for season $t+1$, including players returning from injury. It also
includes any player the team traded away during the summer, since the club could have chosen to keep him. The pool also
contains the team's own free agents, whom the club can re-sign at their actual realized salaries. To evaluate external
additions, the pool incorporates a market of twelve outside free agents and trade targets. This market begins with the
players the team actually acquired that summer. We complete the market with outside movers who logged the most minutes
that year, provided the team could afford them by trade or free agency. Using actual movers at their realized salaries
keeps player availability realistic. The optimization engine can always replicate the moves the front office executed,
but it can also choose differently. A typical pool contains about 24 players.

\subsection{Lineup values from shot patterns}\label{sec:model-value}

To compare potential rosters, a front office must evaluate what five players would produce together, including
combinations that have never shared the court. Historical lineup records cannot answer this question, because
prospective trades and signings create five-player combinations with zero recorded minutes. We therefore construct the
value of any five-player lineup in three steps using data up to season $t$. We focus here on the practical
implications for front-office decisions, and report the full statistical specifications in Appendices~\ref{app:roles}
and~\ref{app:shot}.

\paragraph{Roles.} When evaluating an unfamiliar outside player, a front office cannot measure his chemistry with
existing teammates directly. The front office can, however, determine what kind of player he is, because knowing how a
player type alters teammate shots allows the team to value any newcomer by his role. We describe each player-season by
one of $K=7$ roles using a Bayesian mixture model that groups players with similar box-score statistics and shot charts,
in the tradition of Bayesian clustering of shot charts \citep{hu2021,yin2023}. We chose seven roles for their use to a
front office rather than for forecasting accuracy, because front offices need a small set of recognizable archetypes:
the statistical fit improves little after five or six roles, each role captures a substantial share of league minutes,
and the seventh role separates two distinct forward archetypes that front offices treat differently. Table~\ref{tab:roles} defines the seven roles, and
Figure~\ref{fig:data-players} illustrates a representative player for each role. Rather than assigning a fixed label, the
model gives each player a probability distribution across the seven roles. A player who split his playing time between
creating his own shot and spotting up retains probability mass on both roles, while a player with limited minutes retains
wide uncertainty. The optimization engine draws 20 samples from these posterior distributions as scenarios. Each draw
represents one plausible way player roles could unfold across the league next season.

\begin{table}[tbp]
\centering\small
\caption{The seven roles and their share of league minutes (fit through 2022-23 season).}\label{tab:roles}
\begin{tabularx}{\linewidth}{
    ll
    >{\raggedright\arraybackslash}X
    >{\raggedright\arraybackslash}X
}
\toprule
\textbf{Role} & \textbf{Minutes} & \textbf{How the role plays} & \textbf{Examples}\\
\midrule
Primary creator & 20\% & pull-ups from every zone, high usage, many assists & Edwards, Tatum, LaVine\\
Two-way forward & 18\% & balanced wings and forwards & Mikal Bridges, Scottie Barnes, Harrison Barnes\\
Shooter & 15\% & spot-up and pull-up threes & Caldwell-Pope, Strus, Huerter\\
Spot-up connector & 12\% & corner threes and cuts from low-usage forwards & O'Neale, Finney-Smith\\
Rim big & 12\% & cuts and finishes at the rim, rebounds, blocks & Gobert, Jarrett Allen, Claxton\\
Playmaker & 12\% & drives and assists from guards and point forwards & Smart, Lowry, Draymond Green\\
Hub big & 10\% & post-ups, paint touches and passing from bigs & Joki\'c, Sabonis, Vu\v{c}evi\'c\\
\bottomrule
\end{tabularx}
\end{table}

\paragraph{Shots.} A front office must also account for how incoming players redistribute field-goal attempts across the
roster. Basketball is played with one ball, so adding a second creator takes shots away from players who already create
offense. The shot model predicts how frequently a player shoots from each of the 25 court cells, and how often he
scores, conditional on his on-court teammates. For player $q$ in a stint, defined as an unbroken stretch of play with the
same ten players on the floor, his attempts in cell $c$ follow a Poisson distribution with rate per 100 possessions
\begin{equation}\label{eq:attempts}
A_{qc}=100\exp\big\{\mu_c+s_{qc}+\beta_c^{\top}b_q+(a_q^{\top}u_{g(c)})(b_q^{\top}v_{g(c)})+\gamma\,O_{qc}\big\}.
\end{equation}
The league baseline rate $\mu_c$ and the player's profile $s_{qc}$ describe his baseline attempt frequency beside average
teammates. The vector $b_q$ averages the role probabilities of his four on-court teammates, so $\beta_c^{\top}b_q$
captures how the surrounding mix of roles shifts his attempts in cell $c$. Because teammate role probabilities sum to
one, each coefficient represents a relative shift \citep{li2023}, defined as the change in his attempt rate when an
average teammate is replaced by a teammate of a specific role. The bilinear term
$(a_q^{\top}u_{g(c)})(b_q^{\top}v_{g(c)})$ allows this shift to depend on the shooter's own role $a_q$ through a low-rank
interaction within cell group $g$. This interaction captures asymmetric adjustments: adding a second creator takes shots
from an existing creator, but can generate open shots for a spot-up shooter. The spatial crowding term $\gamma\,O_{qc}$
accounts for congestion that broad roles miss. Here $O_{qc}$ denotes how often his four teammates shoot from that same
cell, so a negative coefficient $\gamma$ means that players who favor identical spots shoot less when sharing the court.
A logistic model with matching teammate covariates estimates whether each attempt is made. Court cells that respond
similarly to teammates share strength through the spatially constrained Bayesian clustering of \citet{meng2025}, the precisions of the teammate effects are chosen by the marginal likelihood on season $t$ itself, and the model is fitted on every stint of the
season (Appendix~\ref{app:shot}). We define two benchmark
models for comparison: shooter-only rates, which retain individual profiles without teammate effects, and additive
roles, which keep teammate roles but drop both the shooter interaction and the crowding term. We refer to formulation
\eqref{eq:attempts} as the full model.

\paragraph{Lineup values.} To compare alternative rosters during summer negotiations, a front office needs a single
performance metric for each five-player lineup. For any candidate lineup $\ell$ and posterior role draw $d$, the shot
model supplies expected attempts and make probabilities for each player across all 25 cells. We multiply expected
attempts by make probabilities and shot point values ($\pts_c$, equal to two or three points), and add expected points
from free throws drawn by each player. This calculation yields the lineup's expected shot points, denoted $\Pi_{\ell d}$,
over $N_{\ell d}$ total attempts per 100 possessions. The engine values each lineup by
\begin{equation}\label{eq:value}
V_{\ell d}=\frac{\Pi_{\ell d}}{N_{\ell d}}\,\min\{N_{\ell d},\bar N\},
\end{equation}
which multiplies points per shot by effective shot volume, capping total attempts at the league average $\bar N$. The
cap reflects how the game works. A lineup that cannot generate shots is charged for the shots it does not take, which
preserves the value of shot creators. Conversely, five high-volume scorers cannot take more than a game's worth of
shots, so stacking them earns nothing extra. Because the underlying model predicts how attempts
shift with teammate composition, this value function directly prices player fit. The same player contributes more to a
lineup that needs his shot creation than to a lineup that already possesses it. We apply two adjustments so that these
valuations travel to rosters that do not exist yet. We damp predicted shot adjustments for newly acquired players by
$\kappa=0.75$, an empirical transfer factor estimated from players who changed teams
(Section~\ref{sec:results-forecast}). We also value any player who missed season $t$ due to injury using his
performance from season $t-1$. With lineup values $V_{\ell d}$ established for all candidate lineups and role draws, the
roster optimization problem is fully specified.

\subsection{The two-stage stochastic program}\label{sec:model-program}

We formulate the offseason decision of Section~\ref{sec:model-problem} as a risk-averse two-stage stochastic
mixed-integer program. Table~\ref{tab:notation} summarizes the mathematical notation.
The candidate pool $\pool$ combines players under contract for season $t+1$ ($\Rt$), the team's own free agents ($\Own$),
and outside market candidates from free agency ($\FA$) and trades ($\Tr$). The binary variable $k_i \in \{0,1\}$
indicates whether player $i \in \pool$ is on the roster for season $t+1$, while $y_i \in \{0,1\}$ indicates whether he
enters the active rotation. The set $\lineups$ contains all feasible candidate lineups of five players from $\pool$ that
maintain positional balance by including at least one guard, at most two bigs, and at most $n^{\mathrm{in}}$ outside
additions. In the second stage, after role uncertainty resolves into scenario $d \in \draws$, the coach selects the
recourse variables $m_{\ell d} \ge 0$, representing the share of team possessions played by lineup $\ell$ in draw $d$.
The team's offensive performance under scenario $d$ is the minutes-weighted value of its active lineups:
\begin{equation}\label{eq:teamvalue}
T_d=\sum_{\ell\in\lineups}V_{\ell d}\,m_{\ell d}.
\end{equation}

\begin{table}[tbp]
\centering\small
\caption{Notation of the two-stage program.}\label{tab:notation}
\begin{tabularx}{\textwidth}{lX}
\toprule
\multicolumn{2}{l}{\emph{Sets}}\\
$\pool=\Rt\cup\Own\cup\FA\cup\Tr$ & the team's pool: players under contract for $t+1$, its own free agents, outside
free agents and trade targets\\
$\FA^{+}\subseteq\FA$ & outside free agents above the minimum salary\\
$G,B\subseteq\pool$ & guards and bigs\\
$\mathcal{A}\subseteq\Rt$ & protected players: All-NBA in season $t$ or $t-1$ and under contract\\
$\lineups$ & candidate lineups: five players of $\pool$, at least one guard, at most two bigs, at most
$n^{\mathrm{in}}$ from $\FA\cup\Tr$\\
$\draws$ & posterior role draws, the scenarios, $|\draws|=D=20$\\
$\cells$ & the 25 cells of the court (zone by movement type)\\
\multicolumn{2}{l}{\emph{Parameters}}\\
$V_{\ell d}$ & value of lineup $\ell$ in draw $d$, points per 100 possessions, from \eqref{eq:value}\\
$S_i$, $S_0$ & next season's salary of player $i$; salaries of the rest of the roster (\$M)\\
$\bar P$ & the owner's payroll limit: the first apron, or the team's actual payroll if higher\\
$n^{\mathrm{in}}, n^{\mathrm{out}}$ & market players brought in and players under contract let go, at most\\
$\underline f,\overline f,\bar m$ & a rotation player's minimum and maximum share of possessions, and a lineup's maximum
(10\%, 75\%, 35\%)\\
$\rho,\alpha$ & weight on the tail term and the tail share (0.5 and 0.2)\\
$\Gamma,\omega$ & cells an opposing defense may attack, and the weight on the stress loss\\
$\lambda_{\ell dc}$ & value lineup $\ell$ loses in draw $d$, per unit of its possession share, when cell $c$ is
attacked\\
Cap, RE & the salary cap and the room exception\\
MLE, MLE$^{\mathrm{tax}}$, BAE & the mid-level, taxpayer mid-level and bi-annual exceptions\\
Apron$_1$, Apron$_2$, Line & the first and second aprons; the line above which trades follow the narrow rule (the tax
line, or the first apron from 2023-24)\\
\multicolumn{2}{l}{\emph{First-stage decisions}}\\
$k_i\in\{0,1\}$ & player $i$ is on the roster for $t+1$ (kept, re-signed or acquired)\\
$y_i\in\{0,1\}$ & player $i$ is in the rotation\\
$z^{\mathrm{room}},z^{\mathrm{hard}},z^{\mathrm{exc}},z^{\mathrm{line}},z^{\mathrm{band}}\in\{0,1\}$ & salary modes:
the team uses cap room; is hard-capped at the first apron; uses an exception under the 2023 agreement; ends under the
line; uses the third trade band\\
$x_i\in\{0,1\}$ & player $i$ moves in the room phase: sent away to clear room ($i\in\Rt$) or absorbed into room
($i\in\Tr$)\\
\multicolumn{2}{l}{\emph{Second-stage decisions and auxiliary variables}}\\
$m_{\ell d}\ge0$ & share of possessions played by lineup $\ell$ in draw $d$\\
$\eta$, $\xi_d\ge0$ & threshold and shortfalls of the tail term\\
$\pi_d,q_{dc}\ge0$ & dual variables of the playoff stress test\\
\bottomrule
\end{tabularx}
\end{table}

\paragraph{Objective.} A front office cannot rely on expected scoring alone. An NBA season provides only one chance to
compete, so a roster that scores well on average but fails when player roles conflict is a risky bet. To balance average performance against downside volatility, we optimize a weighted
combination of expected offensive value, conditional value at risk across unfavorable scenarios, and expected losses
from targeted playoff defenses:
\begin{equation}\label{eq:objective}
\max\;\;(1-\rho)\,\frac1D\sum_{d\in\draws}T_d\;+\;\rho\,\Big(\eta-\frac{1}{\alpha D}\sum_{d\in\draws}\xi_d\Big)
\;-\;\frac{\omega}{D}\sum_{d\in\draws}\mathrm{SL}_d,
\qquad \xi_d\ge\eta-T_d,\;\;\xi_d\ge0\;\;(d\in\draws).
\end{equation}
The second term models downside risk using the linear formulation of conditional value at risk established by
\citet{rockafellar2000}. In this formulation, the auxiliary variable $\eta$ represents the value-at-risk threshold at
quantile $\alpha$, while $\xi_d \ge 0$ measures the performance shortfall below that threshold in scenario $d$. At
optimality, the expression $\eta - \frac{1}{\alpha D}\sum_{d\in\draws}\xi_d$ exactly equals the mean offensive value across
the worst $\alpha$ share of draws. This linear representation eliminates non-smooth minimum operators and keeps the
stochastic program computationally tractable as a mixed-integer linear program. We place equal weight on expected
performance and tail risk by setting $\rho=0.5$, and focus on the lowest quintile of outcomes by setting $\alpha=0.2$,
which evaluates the four worst scenarios among the twenty draws. The third term incorporates the stress loss
$\mathrm{SL}_d$ from defensive countermeasures, which we define below; our primary roster optimizations set $\omega=0$.

\paragraph{Roster and rotation.} A front office must respect strict structural rules governing roster size, positional
depth, transaction volume, and financial capacity. We capture these requirements through the first-stage constraints
\eqref{eq:rotation}--\eqref{eq:payroll}:
\begin{align}
&y_i\le k_i\;\;(i\in\pool),\qquad 8\le\sum_{i\in\pool}y_i\le 10,\qquad \sum_{i\in G}y_i\ge3,\qquad\sum_{i\in B}y_i\ge2,
\label{eq:rotation}\\
&\sum_{i\in\FA\cup\Tr}k_i\le n^{\mathrm{in}},\qquad \sum_{i\in\Rt}(1-k_i)\le n^{\mathrm{out}},\qquad
k_i=1\;\;(i\in\mathcal{A}),\label{eq:moves}\\
&S_0+\sum_{i\in\pool}S_ik_i\le\bar P.\label{eq:payroll}
\end{align}
Constraints \eqref{eq:rotation} define the active playing rotation. The first inequality ensures that a player can enter
the rotation only if he is signed to the roster ($y_i \le k_i$). The remaining bounds require a rotation of eight to ten
players, containing at least three guards ($i \in G$) and at least two bigs ($i \in B$). Constraints \eqref{eq:moves} regulate roster turnover. The first inequality caps outside acquisitions from free
agency and trades at $n^{\mathrm{in}}$, while the second inequality restricts departures of contracted players to at most
$n^{\mathrm{out}}$. Re-signing the team's own free agents does not count against these turnover limits, because the club
already owns their rights. The equality $k_i=1$ protects All-NBA players under contract ($i \in \mathcal{A}$). Because
our shot model evaluates offense, an engine that traded away a franchise player on purely offensive metrics would lack
credibility with front offices. Constraint \eqref{eq:payroll} keeps total team spending within the owner's budget $\bar P$,
summing the salaries of the rest of the roster $S_0$ and the salaries $S_i$ of all retained and acquired players.

\paragraph{Minutes.} During the regular season, coaches share the possessions among combinations of rotation players. In the second stage, the model assigns possession shares to candidate lineups
separately for each posterior draw $d \in \draws$:
\begin{equation}\label{eq:minutes}
\sum_{\ell\in\lineups}m_{\ell d}=1,\qquad \underline f\,y_i\le\sum_{\ell\ni i}m_{\ell d}\le\overline f\,y_i\;\;
(i\in\pool),\qquad 0\le m_{\ell d}\le\bar m\;\;(\ell\in\lineups)\qquad(d\in\draws).
\end{equation}
Constraints \eqref{eq:minutes} govern playing-time allocation across lineups. The first equality ensures that lineup shares
sum to all team possessions in each scenario. The second pair of inequalities links the two decision stages: any player
omitted from the rotation receives zero possessions, while each active rotation player receives between
$\underline f=10\%$ and $\overline f=75\%$ of team playing time. Because constraints \eqref{eq:rotation} require
$y_i \le k_i$, unrostered players receive no minutes, ensuring that a lineup plays only when all five of its members are
on the roster. The final upper bound prevents any single lineup from exceeding $\bar m=35\%$ of total possessions, which
keeps playing-time patterns realistic. Because minutes adapt across draws, the coach can reallocate playing time toward
effective lineups under whichever scenario occurs.

\paragraph{Salary-cap rules.} A front office cannot simply assemble any roster that satisfies its total budget, because
the league's collective bargaining agreement dictates which contract mechanisms a team can legally use. These financial
mechanisms change across collective bargaining agreements, and we encode the rules of the 2011, 2017, and 2023 agreements
\citep{nbacba2011,nbacba2017,nbacba2023}. We model these rules using five binary indicators of the team's operating mode and binary
variables $x_i \in \{0,1\}$ for players involved in cap-room transactions.

An offseason operates in two phases. A team that uses cap space ($z^{\mathrm{room}}=1$) first fits its contracted players,
incumbent free-agent holds, and absorbed trade targets under the salary cap, and then signs outside free agents using
remaining room and the room exception. To create cap space, the front office may send contracted players away
($x_i=1$):
\begin{align}
&x_i\le1-k_i,\;\;x_i\le z^{\mathrm{room}}\;\;(i\in\Rt);\qquad x_j\le k_j,\;\;x_j\le z^{\mathrm{room}}\;\;(j\in\Tr),
\label{eq:roomlink}\\
&S_0+\sum_{i\in\Rt}S_i(1-x_i)+\sum_{j\in\Own}S_jk_j+\sum_{j\in\Tr}S_jx_j\le\mathrm{Cap}+M_1(1-z^{\mathrm{room}}),
\label{eq:room}\\
&S_0+\sum_{i\in\Rt}S_i(1-x_i)+\sum_{j\in\Own}S_jk_j+\sum_{j\in\FA^{+}}S_jk_j+\sum_{j\in\Tr}S_jx_j
\le\mathrm{Cap}+\mathrm{RE}+M_2(1-z^{\mathrm{room}}).\label{eq:roomfa}
\end{align}
Constraints \eqref{eq:roomlink} govern the moves of the cap-room phase. They allow such moves only when the team uses cap
room ($z^{\mathrm{room}}=1$): a player under contract ($i \in \Rt$) can be sent away to clear room only if he leaves the
roster, and a trade target ($j \in \Tr$) can be absorbed into room only if he is acquired. Constraint \eqref{eq:room}
ensures that the players under contract who are not sent away to clear room, the re-signed free agents, the absorbed
trade targets, and the salaries of the rest of the roster $S_0$ fit within the salary cap ($\mathrm{Cap}$). Constraint
\eqref{eq:roomfa} ensures that outside free agents signed above the minimum salary ($j \in \FA^{+}$) fit within the room
exception ($\mathrm{RE}$) and remaining cap room.

A team operating over the salary cap ($z^{\mathrm{room}}=0$) can sign outside free agents above the minimum only through
salary-cap exceptions. Using the full mid-level exception or the bi-annual exception, or acquiring a player through a
sign-and-trade transaction, hard-caps the team at the first apron ($z^{\mathrm{hard}}=1$). Under the 2023 agreement,
spending any exception hard-caps payroll at the second apron ($z^{\mathrm{exc}}=1$; constraints \eqref{eq:apron2} apply to
that agreement only):
\begin{align}
&\sum_{j\in\FA^{+}}S_jk_j\le\mathrm{MLE}^{\mathrm{tax}}+(\mathrm{MLE}-\mathrm{MLE}^{\mathrm{tax}}+\mathrm{BAE})\,
z^{\mathrm{hard}}+M_3z^{\mathrm{room}},\label{eq:exceptions}\\
&S_0+\sum_{i\in\pool}S_ik_i\le\mathrm{Apron}_1+M_4(1-z^{\mathrm{hard}}),\qquad k_j\le z^{\mathrm{hard}}\;\;
(j\text{ arriving by sign-and-trade}),\label{eq:hardcap}\\
&\sum_{j\in\FA^{+}}S_jk_j\le(\mathrm{MLE}+\mathrm{BAE})\,z^{\mathrm{exc}}+M_5z^{\mathrm{room}},\qquad
S_0+\sum_{i\in\pool}S_ik_i\le\mathrm{Apron}_2+M_6(1-z^{\mathrm{exc}}).\label{eq:apron2}
\end{align}
Constraints \eqref{eq:exceptions} ensure that an over-the-cap team restricts outside free-agent signings to the taxpayer
mid-level exception ($\mathrm{MLE}^{\mathrm{tax}}$) unless it accepts a hard cap at the first apron ($z^{\mathrm{hard}}=1$).
Constraints \eqref{eq:hardcap} enforce the first apron hard cap. The first inequality ensures that payroll does not exceed
the first apron threshold ($\mathrm{Apron}_1$) when hard-capped, while the second inequality ensures that any player
arriving via sign-and-trade triggers this hard cap. Constraints \eqref{eq:apron2} enforce the second apron rules of the
2023 agreement. The first inequality restricts exception spending unless the team accepts a second apron restriction
($z^{\mathrm{exc}}=1$), and the second inequality ensures that total payroll remains below the second apron
($\mathrm{Apron}_2$).

Trades outside cap room must balance outgoing and incoming player salaries. Let $I=\sum_{j\in\Tr}S_j(k_j-x_j)$ denote
incoming trade salary and $O=\sum_{i\in\Rt}S_i(1-k_i-x_i)$ denote outgoing salary matched in the trade phase. Teams ending
above the salary line must match trades tightly, whereas teams finishing below the line ($z^{\mathrm{line}}=1$) choose
between two flexible trade bands using binary selector $z^{\mathrm{band}}$:
\begin{align}
&S_0+\sum_{i\in\pool}S_ik_i\le\mathrm{Line}+M_7(1-z^{\mathrm{line}}),\qquad
I\le a_0O+b_0+M_8z^{\mathrm{line}},\label{eq:line}\\
&I\le a_1O+b_1+M_9(1-z^{\mathrm{line}}),\qquad I\le O+b_2+M_{10}(1-z^{\mathrm{line}}+z^{\mathrm{band}}),\nonumber\\
&I\le a_3O+b_3+M_{11}(2-z^{\mathrm{line}}-z^{\mathrm{band}}).\label{eq:trades}
\end{align}
Constraints \eqref{eq:line} define operating status relative to the salary line. The first inequality ensures that a
team counted as under the line ($z^{\mathrm{line}}=1$) does finish below it, while the second inequality ensures that a
team ending above the line obeys the narrow trade-matching rule $I \le a_0O + b_0$. Constraints \eqref{eq:trades} govern trades for teams
finishing below the line ($z^{\mathrm{line}}=1$). The first inequality enforces a common upper bound $I \le a_1O + b_1$,
while the remaining inequalities permit the team to choose between the trade bands $I \le O + b_2$ and $I \le a_3O + b_3$
using binary variable $z^{\mathrm{band}}$.

The big-M parameters $M_1$ through $M_{11}$ enable these logical conditions within a linear framework. We calculate each
constant $M_j$ as the maximum amount by which the left side of its constraint can exceed the right side under the owner's
payroll limit $\bar P$. Choosing the smallest valid constants switches off the rules that do not apply in a mode while
keeping the linear relaxation tight. We match trades in aggregate across the summer on the team's side only, and allow minimum-salary signings and incumbent free
agents without matching restrictions. Appendix~\ref{app:rules} provides the numerical values for parameters
$a_0, a_1, a_3$ and $b_0, b_1, b_2, b_3$, as well as the aggregation rules governing multi-player trades from 2024-25.

\paragraph{Protection against playoff defenses.} A front office that optimizes purely for regular-season scoring can build
a roster vulnerable to postseason game plans. Against the best playoff defenses, teams lose a sizeable share of their
cuts to the rim and their spot-up shots in the paint, and take many more pull-up threes instead
(Appendix~\ref{app:stress}). A roster that relies on a few kinds of shot is exposed to this. To guard against this
vulnerability, our formulation provides an optional defensive stress test that allows an opposing defense to remove part
of the team's attempts in up to $\Gamma$ of the 25 court cells. The removed attempts are re-routed to alternative shots
at the lower value they keep in the playoffs. Under scenario $d$, the loss in
cell $c$ depends linearly on playing-time shares, $L_{dc}=\sum_\ell\lambda_{\ell dc}m_{\ell d}$.

The opposing defense chooses an optimal attack by targeting the $\Gamma$ cells that cause the largest reduction in
offensive value. This adversarial attack forms a budgeted uncertainty set \citep{bertsimas2004}. Because a maximization
nested inside the program would make it nonlinear, we replace the defensive subproblem by its linear programming dual:
\begin{equation}\label{eq:stress}
\mathrm{SL}_d=\max_{C'\subseteq\cells,\;|C'|\le\Gamma}\;\sum_{c\in C'}L_{dc}
=\min_{\pi_d,\,q_{dc}\ge0}\Big\{\Gamma\pi_d+\sum_{c\in\cells}q_{dc}\;:\;q_{dc}+\pi_d\ge L_{dc}\;\;(c\in\cells)\Big\}.
\end{equation}
In this dual representation, the variable $\pi_d \ge 0$ sets a baseline threshold on defensive damage, while $q_{dc} \ge 0$
captures cell losses that exceed this threshold. Inserting this minimization into objective \eqref{eq:objective} with
$\omega=1$ keeps the overall program linear and computationally tractable. By varying $\Gamma$ from one to four cells, a
front office can trace, for its own roster, the trade-off between regular-season value and protection against a playoff
defense. Historical
playoff series through season $t$ provide the empirical estimates for shot reductions and re-routed scoring values.

\paragraph{Properties.} The formulation possesses two attractive structural properties. All objective terms and
operational constraints are linear, so the model is a mixed-integer linear program. The formulation is also exact for the
underlying value model. Teammates influence individual scoring only through the five-player lineup on the floor, and
precomputing all candidate lineup values $V_{\ell d}$ avoids inner approximations. To isolate what drives decision
quality, we define two benchmark formulations. The mean-roles program condenses the twenty draws into a single scenario
where each player receives his expected role probabilities. The single-stage program fixes playing time before role
uncertainty resolves, requiring $m_{\ell d}=m_\ell$ across all draws.

\subsection{Solution method and evaluation}\label{sec:model-solve}

Solving the two-stage program requires an efficient solution strategy to handle the large number of continuous recourse
variables. A typical pool of about 24 players generates some 19,000 candidate lineups, so the twenty-draw program
contains several hundred thousand continuous playing-time variables alongside fewer than one hundred binary roster
variables. We precompute all lineup values using JAX \citep{bradbury2018} on CPU or GPU, which takes under a minute per team-offseason across all models and draws. We construct each problem instance as a sparse matrix and solve it with HiGHS
\citep{huangfu2018} to a relative optimality gap of $10^{-4}$, setting a time limit of one hour per program.

When initialized without a starting solution, the solver can search for an hour without discovering an effective roster.
We resolve this difficulty with a sequential three-step procedure. We first solve the smaller mean-roles program to
optimality. We then fix that roster and optimize rotation and playing-time variables across all twenty draws, which
supplies an advanced feasible solution for the full program. We initialize the twenty-draw program from this solution,
disabling solver presolve because presolve alone requires about twenty minutes on instances of this scale.

This solution procedure allows the engine to achieve a relative optimality gap of $10^{-4}$ in every backtest case,
guaranteeing an objective value that is never inferior to the mean-roles plan under flexible minutes.
Table~\ref{tab:computation} reports problem sizes and computation times. Solving all programs for one team-offseason takes
about 25 minutes on a standard server, and the backtest ran across five servers sharing a common task queue.

\begin{table}[tbp]
\centering\small
\caption{Problem sizes and solver statistics over the 240 team-offseasons (engine with the full model).}
\label{tab:computation}
\begin{tabular}{lccc}
\toprule
& Median & 90th percentile & Maximum\\
\midrule
Players in the pool & 24 & 26 & 27\\
Candidate lineups & 18,650 & 32,900 & 49,390\\
Relative gap of the recourse program & $2.5\times10^{-5}$ & $9.1\times10^{-5}$ & $1.0\times10^{-4}$\\
All programs of one team-offseason (minutes) & 24 & 62 & 187\\
\bottomrule
\end{tabular}
\end{table}

We verified the mathematical formulation through two independent checks. On 13 instances small enough to permit complete roster enumeration, under the 2011, 2017, and 2023 agreements and six of them signing free agents into cap room,
the program matched the brute-force optimum in every case. Every roster recommended by the engine in the backtest also
satisfies an independent audit of league salary rules. The actual rosters assembled by teams pass that same audit in 74\%
of team-offseasons.

\paragraph{How the plans are judged.} Beyond evaluating each recommended plan, a front office needs to know which model
components drive the suggested decisions. For any roster $k$, we denote by $\hat z(k)$ its value under objective
\eqref{eq:objective} and the full model, with playing-time recourse variables re-optimized in every draw. By evaluating
all plans under $\hat z$, we judge plans produced by simpler models by what they would deliver if the full model is
right. We structure our evaluation around four primary comparisons that directly address the research
questions of Section~\ref{sec:questions}:
\begin{itemize}
\item the \emph{gain over the real roster}, $\hat z(k^{\star})-\hat z(k^{\mathrm{real}})$, where $k^{\star}$ is the
engine's plan and $k^{\mathrm{real}}$ the roster the team fielded, with minutes optimized over its own lineups;
\item the \emph{value of the stochastic solution} \citep{birge2011}, $\VSS=\hat z(k^{\star})-\hat z(k^{\mathrm{mean}})$,
the gain from planning for the uncertainty about roles rather than for each player's average role, where
$k^{\mathrm{mean}}$ is the roster of the mean-roles program;
\item the \emph{value of modeling synergy}, $\hat z(k^{\star})-\hat z(k^{\mathrm{add}})$ and
$\hat z(k^{\star})-\hat z(k^{\mathrm{shoot}})$, where $k^{\mathrm{add}}$ and $k^{\mathrm{shoot}}$ are the plans the engine
makes when lineup values come from additive roles or from shooter-only rates;
\item the value of the crowding term, measured in the same way against the plan made without it, and the trade-off
between regular-season value and playoff protection traced by the stress test.
\end{itemize}
We evaluate the value of flexible minutes by comparing the recourse program with its single-stage counterpart. Each
comparison evaluates decision quality rather than statistical forecasting. A richer model earns its place only if the
plans it produces differ from, and score better than, the plans of a simpler one.

\section{Results}\label{sec:results}

This section answers the first three research questions raised in Section~\ref{sec:questions}; Section~\ref{sec:cases} addresses the fourth. We evaluate every decision model and statistical forecast by fitting the models on data up to season $t$ and evaluating them on season $t+1$ across the eight consecutive season pairs. Nothing from the later season enters any estimate, so every evaluation is out of sample.

\subsection{Teammate fit can be predicted}\label{sec:results-forecast}

\begin{figure}[tbp]
\centering
\includegraphics[width=0.8\textwidth]{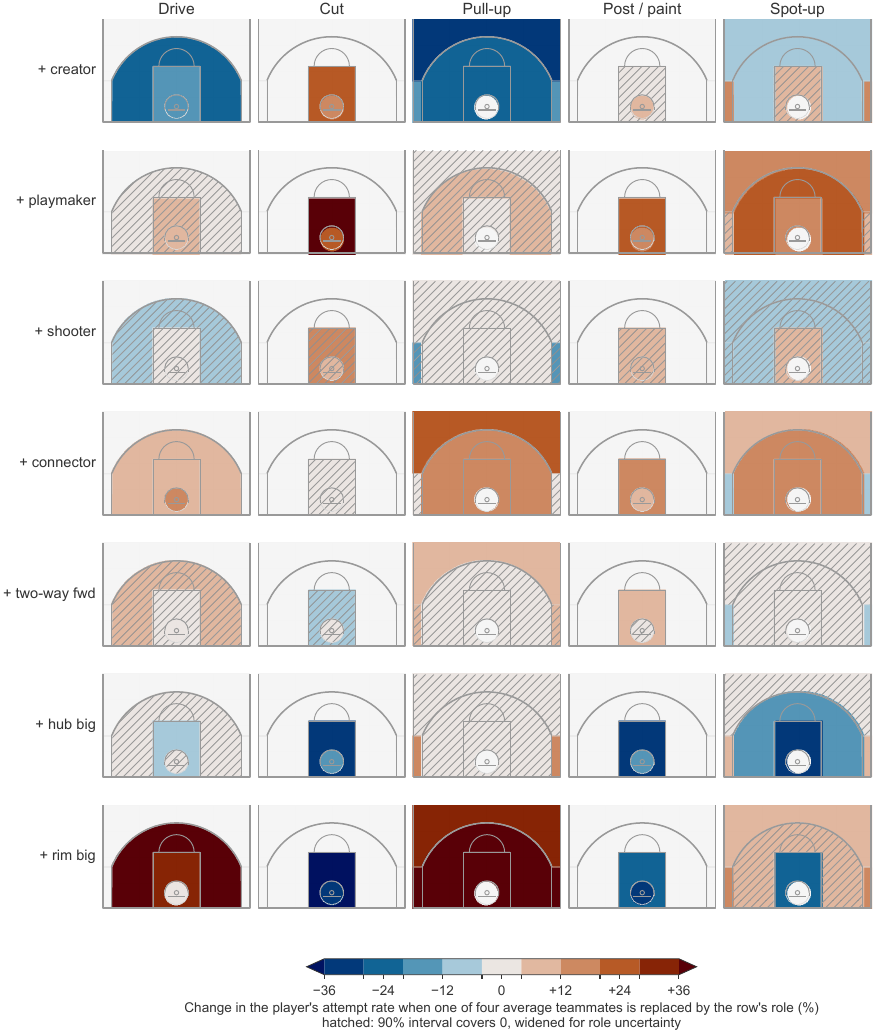}
\caption{How one teammate's role changes a player's shots (2022-23). Each row replaces one of a player's four average
teammates by a teammate of that role; each panel is a movement type drawn over the five court zones, and the color is
the change in the player's attempt rate. Hatched cells: the 90\% interval includes zero once the uncertainty about roles
is taken into account.}\label{fig:effects}
\end{figure}

This subsection answers Question Q1 by establishing that teammate fit can be predicted from micro-level shot data. The estimated shot model captures the concrete tactical trade-offs coaches see on the court (Figure~\ref{fig:effects}). When a primary creator joins a lineup in place of an average teammate, a player's above-the-break pull-ups fall by almost a third, because the new creator takes those shots himself. When a playmaker joins, the player's cuts to the rim rise by a quarter, because someone is now there to find him. A rim big shifts the player's shot distribution away from the basket, reducing cuts and post-ups while increasing drives and pull-ups. These adjustments reflect the choices front offices weigh when pairing scorers with creators, playmakers, or traditional centers, and the model puts a number on each shift.

\begin{figure}[tbp]
\centering
\includegraphics[width=0.9\textwidth]{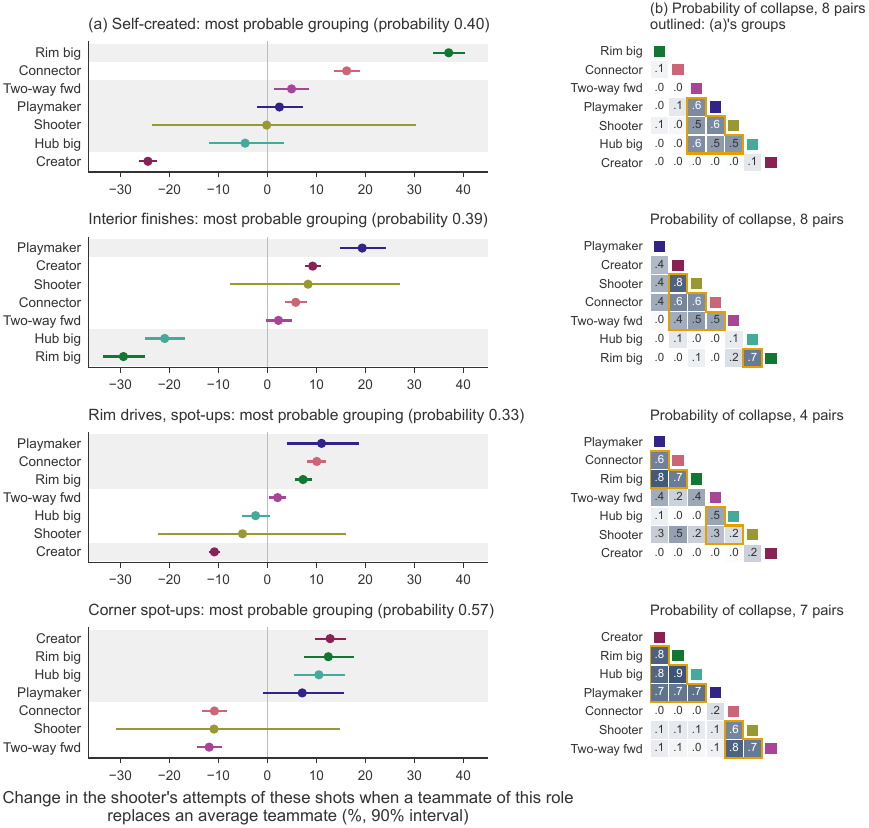}
\caption{Which teammate roles act alike, by kind of shot. (a) The role effects on the shots of each cell cluster in the
latest season pair (replacing an average teammate, with 90\% intervals), with the roles ordered and shaded by the most
probable grouping. (b) The probability that two roles have equal effects, averaged over the season pairs that have the
cluster, with the groups of (a) outlined.}\label{fig:collapse}
\end{figure}

Court geography and teammate roles exhibit stable structure that simplifies front-office decisions. Cells that respond to teammates in the same way form groups that recur season after season: self-created shots (pull-ups and drives away from the rim), finishes near the basket (cuts and post-ups), and, usually, corner spot-ups (Appendix~\ref{app:shot}). Within each group, some teammate roles have the same effect. Following the collapsed relative-shift patterns of \citet{xu2026}, a posterior over the 877 groupings of the seven teammate roles within each cell cluster shows which roles have equal effects (Figure~\ref{fig:collapse}). In most seasons, for corner spot-ups, the spacing roles (spot-up connector, two-way forward, shooter) cut a player's corner threes and the other four raise them. For self-created shots, the primary creator, the rim big, and the spot-up connector each stand apart. Roles that act alike on a kind of shot are interchangeable in their effect on it, which narrows the comparisons a front office has to make.

Out-of-sample evaluations confirm that these teammate effects predict shooting patterns in the subsequent season. Table~\ref{tab:ladder} measures forecasts of the next season's attempts by their log score \citep{gneiting2007}, against shooter-only rates. Knowing a player's teammates improves the forecast of his shots substantially, in every season, and teammates' roles bring by far the largest gain. Grouping similar cells preserves predictive accuracy while making the effects readable. Letting effects depend on the player's own role adds a modest improvement, and adding the crowding term provides a clear further gain across all eight season pairs.

\begin{table}[tbp]
\centering\small
\caption{How each part of the shot model improves the forecast of the next season's shots: gain in log score per game
over shooter-only rates, averaged over the eight season pairs, and the number of pairs in which the step itself improves
the forecast.}\label{tab:ladder}
\begin{tabular}{lcc}
\toprule
Model & Gain & Pairs improved\\
\midrule
+ teammates' roles, each cell on its own & $+2.05$ & 8 of 8\\
+ similar cells grouped (\emph{additive roles}) & $+2.06$ & 5 of 8\\
+ effects that depend on the player's own role & $+2.13$ & 5 of 8\\
+ crowding beyond roles (\emph{the full model}) & $+2.35$ & 8 of 8\\
\bottomrule
\end{tabular}
\end{table}

\begin{figure}[tbp]
\centering
\includegraphics[width=\linewidth]{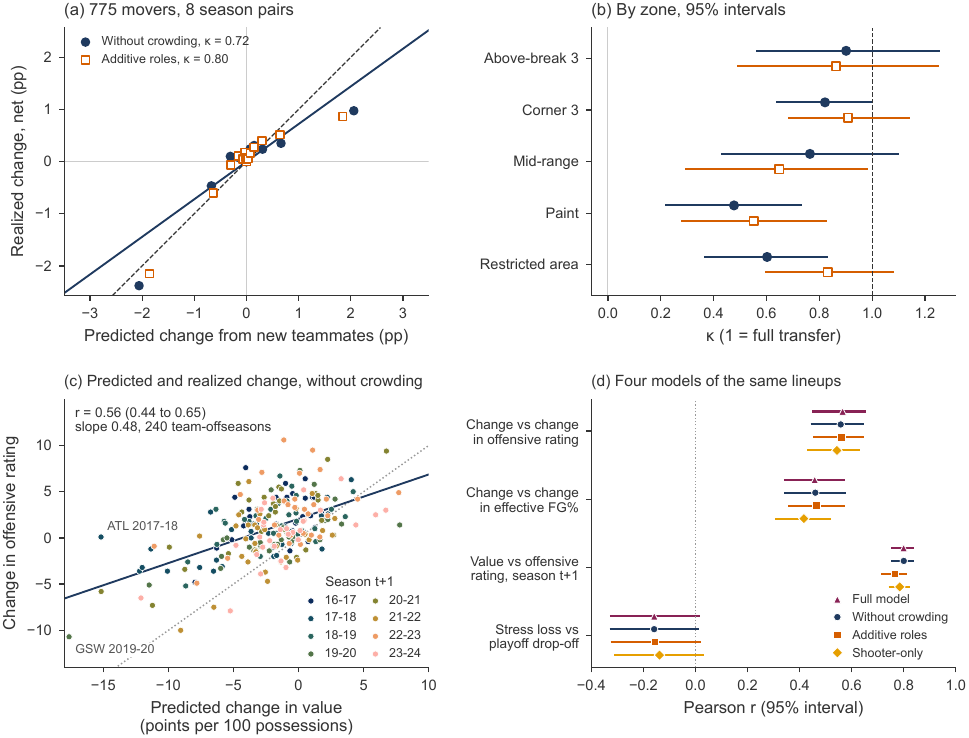}
\caption{The effects hold on the next season. (a) Players who change teams: the realized change in a mover's shot mix
against the change the model predicts from his new teammates, in bins; the dashed line is full transfer. (b) The
transfer factor $\kappa$ by zone, with 95\% intervals. (c) Lineup values and team offense, 240 team-seasons: the
predicted change in the value of the lineups a team played against the realized change in its offensive rating, under
the model without the crowding term; the solid line is the fitted line. (d) Correlations with 95\% intervals for four
versions of the model.}\label{fig:forecast}
\end{figure}

The predicted teammate effects carry over when players change teams. Because the engine values candidates on new rosters, the effects must hold when a player moves. For the 775 players who played substantial minutes for one team and then for another, we compare the change in their shot mix with the change the model attributes to their new teammates (Figure~\ref{fig:forecast}a,b). The predicted effects carry over at about three quarters of their size, with weaker transfer near the basket than for threes. We therefore apply this transfer factor $\kappa=0.75$ as a damping parameter whenever the engine evaluates newcomers.

Lineup values derived from shot patterns directly track realized team scoring on the court. For each team, we value the lineups it played in two consecutive seasons, using only data up to the first season. When the value of a team's lineups rises, its offensive rating rises too (Figure~\ref{fig:forecast}c), and in levels the value of the lineups a team uses closely tracks how well it scores (panel d). The lineup values also carry predictive signal that past records omit: they forecast next season's offense better than last season's offensive rating does, both in levels and in changes. Adding them to last season's rating raises the share of explained variation from 43\% to 67\%. A point of predicted change goes with about half a point of realized change, so the values are best read as a ranking of rosters whose differences are somewhat exaggerated; we use this calibration when we translate the engine's gains into offensive rating. The answer to Question Q1 is therefore yes: teammates' roles shift where and how often players shoot, these shifts carry over to players who change teams, and the resulting lineup values track how well teams score.

\subsection{Better rosters than the ones teams built}\label{sec:results-rosters}

This subsection answers the first part of Question Q2: we test whether planning for teammate fit builds better rosters than the ones front offices actually assembled across 240 historical team-offseasons. We put the engine in the position of every front office in every summer from 2016 to 2023, thirty teams over eight summers. Each problem instance uses only what the team could have known at the time: the models estimated on data up to season $t$, the team's pool and market, the salary rules of the coming league year, and the twenty posterior role draws. To keep the cases comparable, every team may bring in two market players ($n^{\mathrm{in}}=2$, more if it needs them to field a rotation) and let three players under contract go ($n^{\mathrm{out}}=3$). The engine then plans the offseason, and we compare its plan with the roster the team actually assembled through its real trades, signings, and departures. For the comparisons of Section~\ref{sec:model-solve}, the engine also plans each case with the simpler models, and every plan, including the real roster, is scored in the same way.

The engine identifies rosters with substantially higher predicted offensive value than those assembled by front offices in practice. In almost every case, 99\% across the backtest, the engine finds a roster that the model values more highly than the one assembled by the team, gaining about five and a half points per 100 possessions on average (Table~\ref{tab:summers}). The gap is widest in the summers of 2016 and 2017 and settles around four points from 2019 on. Because a point of predicted gain goes with about half a point of realized gain (Section~\ref{sec:results-forecast}), the average gap corresponds to about 2.8 points of offensive rating in real play. That improvement is large enough to lift a league-average offense to about the sixth best in the league.

\begin{table}[tbp]
\centering\small
\caption{The backtest by summer (30 teams each; points per 100 possessions). Gain: the engine's plan over the real
roster. The last four columns give what a front office would lose by planning without each part of the
model.}\label{tab:summers}
\begin{tabular}{lcccccc}
\toprule
& \multicolumn{2}{c}{Engine against the real roster} & \multicolumn{4}{c}{Value of planning with the full model}\\
\cmidrule(lr){2-3}\cmidrule(lr){4-7}
Summer & Gain & Engine better & Role uncertainty & vs additive & vs shooter-only & Crowding\\
\midrule
2016 & 8.04 & 100\% & 0.18 & 0.41 & 1.70 & 0.01\\
2017 & 10.54 & 100\% & 0.16 & 0.77 & 1.99 & 0.03\\
2018 & 5.77 & 100\% & 0.19 & 0.35 & 1.26 & 0.04\\
2019 & 4.07 & 97\% & 0.14 & 0.66 & 1.01 & 0.03\\
2020 & 4.04 & 100\% & 0.27 & 0.53 & 1.84 & 0.02\\
2021 & 4.65 & 100\% & 0.45 & 0.48 & 0.96 & 0.09\\
2022 & 4.46 & 100\% & 0.10 & 0.75 & 1.15 & 0.04\\
2023 & 3.67 & 97\% & 0.17 & 0.43 & 1.11 & 0.05\\
\midrule
All 240 & 5.65 & 99\% & 0.21 & 0.55 & 1.38 & 0.04\\
\bottomrule
\end{tabular}
\end{table}

The engine's plans translate the optimization results into salary-compliant rosters through feasible player transactions. They bring in two market players and send out two or three players on average, and every plan satisfies the salary rules. Many of them create cap space by sending away players under contract, which the market of a real summer would have to absorb. The engine is not simply buying stars: since the market contains the players each team really acquired, the engine could often have repeated the team's moves, and it almost always found a better combination. In answer to the historical part of Question Q2, planning for teammate fit within salary constraints builds substantially better rosters than real NBA front offices constructed.

\subsection{What drives the decision}\label{sec:results-drivers}

\begin{figure}[tbp]
\centering
\includegraphics[width=\linewidth]{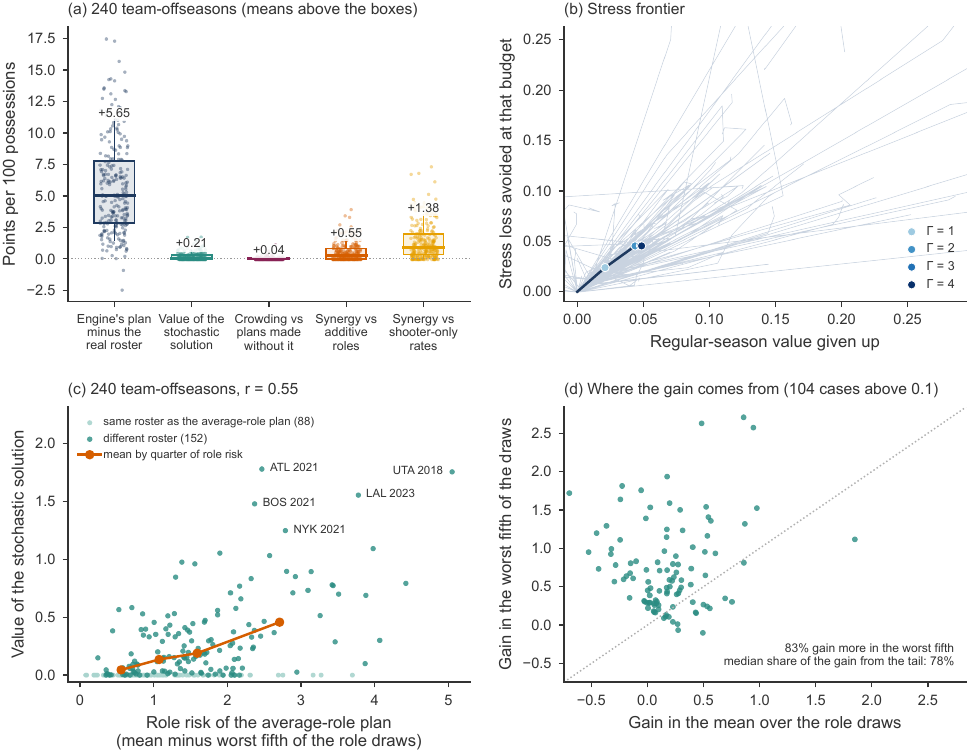}
\caption{What drives the decision, over 240 team-offseasons (points per 100 possessions). (a) The engine's plan against
the real roster, and what is lost by planning with a simpler model: with each player's average role, without the
crowding term, with additive roles, and with shooter-only rates. Every plan is scored by the full model; means are
printed above the boxes. (b) Protection against a playoff defense: plans made for an attack on $\Gamma$ cells give up
regular-season value (horizontal) to avoid losses against that attack (vertical); thin lines are teams, dots are means.
(c) The value of the stochastic solution against the risk of the average-role plan (its mean value minus the mean of its
worst fifth of draws); light points are cases where both plans pick the same roster, and the orange line follows the
mean by quarter of risk. (d) For the cases with a gain above 0.1, the gain in the average over the draws against the
gain in the worst fifth of the draws.}\label{fig:decisions}
\end{figure}

This subsection answers Question Q3 by determining which components of the optimization model drive roster choices and quantifying what a front office loses by omitting them. Figure~\ref{fig:decisions}(a) establishes this by showing what a front office loses when it plans with less of the model. The clearest answer concerns teammate synergy. A front office that valued each player by his own shooting, as if his teammates did not matter, would choose the engine's roster in only 7\% of the cases, and its rosters would score well over a point per 100 possessions less. The reason is the competition for shots of Section~\ref{sec:model-value}. Judged one at a time, two scorers who take the same shots look as good together as apart; once their competition is counted, they are worth less together than a scorer paired with a playmaker or a shooter. A front office that thinks in roles but treats their effects as simply additive gets part of the way, keeping the engine's roster in about one case in five and losing about half a point. Fit, in short, is not a refinement of player value: it changes which players a team should pursue.

Teammate effects also govern which trades make economic sense for both franchises. Holding each team's real roster for the next season, we screened every one-for-one and two-for-two trade of rotation players within the salary rules of both teams and solved the most promising of them exactly, up to 200 a season. Under the full model, between 0 and 28 trades a season raise both teams' value in all 20 role draws (94 over the eight seasons), against 53 to 193 under shooter-only rates. Without teammate effects, a trade can help both sides only through the cap on a lineup's attempts, while the full model charges an incoming creator for the shots he takes from his new teammates.

The remaining model components exert much smaller effects on roster selection. With the crowding term, the engine picks a different roster in about half of the cases, but the rosters it would otherwise have picked are almost as good: the term helps the engine choose among rosters that are already close in value. Protection against a strong playoff defense is cheap. A roster built to withstand an opponent who cuts into two of its favorite kinds of shot gives up only a few hundredths of a point per 100 possessions in the regular season (Figure~\ref{fig:decisions}b), and larger attacks add almost nothing.

Uncertainty about player roles does not require elaborate hedging because in-season playing-time adjustments absorb most risk. Front offices do not know in advance whether a newcomer will create his own shots or spot up. In most cases, however, planning for every role draw yields little gain over planning for each player's average role and letting rotation minutes adapt. The value of the stochastic solution averages 0.21 points (Table~\ref{tab:summers}), and in more than a third of the cases both formulations pick exactly the same roster. The second stage performs this heavy lifting: whatever roles the season reveals, the coach moves minutes toward lineups that click, so a roster built for average roles rarely leaves value behind.

Stochastic planning pays off when an offseason roster carries high role volatility. Figure~\ref{fig:decisions}(c) shows that the gain from stochastic modeling grows with the risk of the average-role plan, measured by how far its value drops in its worst draws. For the safest quarter of rosters, the value of the stochastic solution is close to zero; for the riskiest quarter, it averages about half a point per 100 possessions, and for a few teams it is well over a point. Most of this gain comes from the bad draws (panel d): planning for uncertainty changes little in average outcomes, but protects the franchise against seasons where new roles combine poorly. The cases where it matters most cluster in the summer of 2021, perhaps because the shortened 2020-21 season left many players' roles less settled than usual.

Front offices typically have multiple near-optimal rosters available at similar salaries. Table~\ref{tab:variants} repeats the backtest with the engine without the crowding term under three changes to its inputs. When newcomers' effects are damped more or less ($\kappa=0.5$ or $\kappa=1$), the plan changes in about half of the cases while the value it reaches barely moves. When the free agents who re-signed with their own teams are removed from the market, most plans change but the engine's lead over the real rosters stays almost the same. This version of the engine reaches the same conclusions as the full one: its lead over the real roster is 5.66 points (median 4.86), the value of its stochastic solution 0.29, its values of modeling synergy 0.52 against additive plans and 1.69 against shooter-only plans, and its stress frontier gives up 0.04 to save 0.05 at $\Gamma=2$. Within salary constraints, a team's market usually offers several players who satisfy the same positional need at comparable cost. The engine's most useful output is therefore a short list of strong rosters, among which managers can choose using attributes outside the model, such as defense, health, and character. In answer to Question Q3, teammate fit is the primary driver of roster decisions: flexible minutes absorb most of the uncertainty about roles, protection against playoff defenses costs little, and the engine offers front offices a short list of near-optimal rosters rather than a single answer.

\begin{table}[tbp]
\centering\small
\caption{Robustness of the plans, 240 team-offseasons (239 without re-signers: one case has no feasible roster), engine
without the crowding term. Points per 100 possessions; mean (median). Change: the attainable value minus the main
run's.}\label{tab:variants}
\begin{tabular}{lcccc}
\toprule
Variant & Change & Same plan & Engine minus real & VSS\\
\midrule
Main run ($\kappa=0.75$) & --- & --- & 5.66 (4.86) & 0.29 (0.07)\\
$\kappa=0.5$ & $+0.12$ ($+0.04$) & 50\% & 5.78 (4.89) & 0.25 (0.07)\\
$\kappa=1$ & $+0.03$ ($+0.06$) & 58\% & 5.60 (4.98) & 0.32 (0.11)\\
No other teams' re-signers & $-0.32$ ($-0.00$) & 31\% & 5.36 (4.76) & 0.36 (0.11)\\
\bottomrule
\end{tabular}
\end{table}

\subsection{Two held-out seasons}\label{sec:results-sealed}

This subsection answers the second part of Question Q2 by evaluating whether the gains from planning for fit hold out of sample in two seasons kept aside during model development. All results discussed so far derive from seasons studied while refining the methodology. To ensure that our findings are not an artifact of model tuning, we held two complete seasons untouched. Nothing from 2024-25 was inspected, and data from 2025-26 were not even downloaded, until the code was frozen at a single commit and ten claims were written down, each with the range it took over the eight earlier seasons. Each untouched season was then processed and scored with the models estimated on the season before it. Three data problems that surfaced during these runs were fixed by correcting how the data are read, without changing any model, setting or claim, and every affected case was solved again (Appendix~\ref{app:protocol}).

The central claims held in both held-out seasons (Table~\ref{tab:sealed}). Teammates' roles improved shot forecasts by amounts comparable to the strongest historical results; the estimated shifts carried over to players changing teams; lineup values tracked realized team offense; the engine constructed a better roster than the real one for every team; and plans made without teammate effects sacrificed value. The crowding term again improved forecast accuracy, while the interaction with a player's own role added no predictive power and degraded forecast accuracy in 2024-25. The answer to Question Q2 therefore holds on seasons the method had never seen: planning for teammate fit builds better rosters than the ones teams assembled.

\begin{table}[tbp]
\centering\small
\caption{The claims on the two held-out seasons, beside their range over the eight earlier seasons. Claims 1--3: gain in
log score per game; claim 4: the transfer factor $\kappa$; claims 5 and 6: correlations over 30 teams; claims 7--10:
points per 100 possessions.}\label{tab:sealed}
\begin{tabularx}{\textwidth}{lXccc}
\toprule
& Claim & Earlier seasons & 2024-25 & 2025-26\\
\midrule
1 & Teammates' roles improve forecasts of players' shots & $+1.43$ to $+2.60$ & $+2.51$ & $+2.68$\\
2 & The dependence on a player's own role adds little & $-0.09$ to $+0.26$ & $-0.65$ & $-0.02$\\
3 & Players shoot less where their teammates shoot & $+0.13$ to $+0.35$ & $+0.05$ & $+0.07$\\
4 & Teammate effects carry over to players who move & 0.44 to 1.21 & 0.50 & 0.93\\
5 & Lineup values track changes in team offense & 0.36 to 0.73 & 0.67 & 0.55\\
6 & Lineup values rank team offenses & 0.54 to 0.84 & 0.79 & 0.69\\
7 & The engine improves on the real roster & 3.67 to 10.54 & 4.50 & 3.83\\
8 & Planning for role uncertainty adds a little & 0.10 to 0.45 & 0.07 & 0.14\\
9 & Teammate effects change the plan: vs additive roles & 0.35 to 0.77 & 0.58 & 0.82\\
& \quad vs shooter-only rates & 0.96 to 1.99 & 0.91 & 1.49\\
10 & Crowding changes plans more than values & 0.01 to 0.09 & 0.10 & 0.07\\
\bottomrule
\end{tabularx}
\end{table}

\section{Case studies: three famous summers}\label{sec:cases}

\begin{figure}[tbp]
\centering
\includegraphics[width=0.88\textwidth]{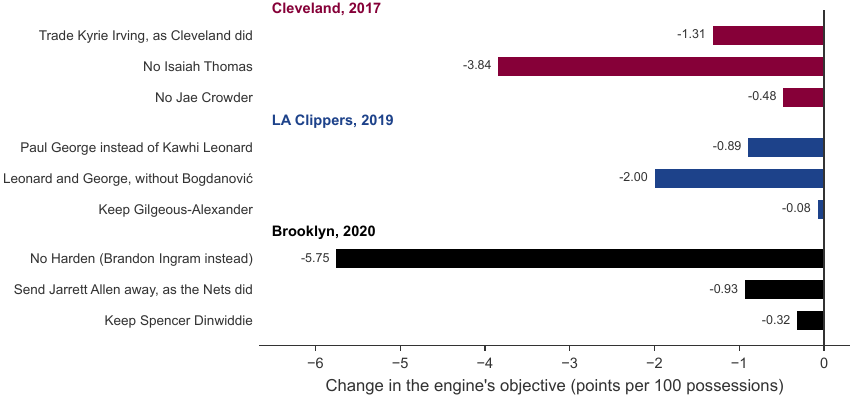}
\caption{What the key moves of three summers are worth for each team. Each bar is the change in the engine's objective
when the program is solved again with one move imposed or ruled out (points per 100 possessions).}\label{fig:moves}
\end{figure}

This section answers Question Q4: we examine how the engine evaluates high-profile front-office decisions that shaped the league, following Cleveland in 2017, the Los Angeles Clippers in 2019, Brooklyn in 2020, and Boston and Phoenix in 2023. While the broad backtest demonstrates systematic gains across 240 team-offseasons, these case studies illuminate the concrete basketball reasoning of the stochastic program. To inspect each decision in realistic operational detail, we let each team make at most as many moves as the franchise made that summer, rather than applying the uniform limits of the backtest. We also price individual transactions by re-solving the stochastic program with specific players forced onto the roster or ruled out, using identical lineup values and posterior draws (Figure~\ref{fig:moves} and Table~\ref{tab:moves}); unchanged re-solves reproduce the original objective values. These are the numbers a front office would want before it commits: what a star is worth to this particular roster, what the best alternative would have been, and what a concession in a trade costs.

\begin{table}[tbp]
\centering\footnotesize
\caption{Single-move re-solves of the three case studies (engine with the full model). Change: the objective minus the
case's, points per 100 possessions. Re-signings of the team's own free agents are omitted.}\label{tab:moves}
\begin{tabularx}{\textwidth}{llcXX}
\toprule
Case & Move & Change & Acquired & Sent away\\
\midrule
Cleveland, 2017 & engine's plan (122.23) & --- & Thomas, Crowder, Rose & Smith\\
& Irving sent away & $-1.31$ & Thomas, Crowder, Rose, Wade & Irving\\
& Thomas ruled out & $-3.84$ & George, Crowder, Rose, Wade & Love\\
& Crowder ruled out & $-0.48$ & Thomas, George & Love\\
\midrule
LA Clippers, 2019 & engine's plan (126.70) & --- & Leonard, Bogdanovi\'c, Noah & Gilgeous-Alexander, Robinson\\
& George forced in & $-0.89$ & George, Bogdanovi\'c, Noah & Gilgeous-Alexander, Robinson\\
& Leonard ruled out & $-0.89$ & George, Bogdanovi\'c, Noah, Patterson & Gilgeous-Alexander, Robinson\\
& Bogdanovi\'c ruled out & $-2.00$ & Leonard, George, Noah & L.\ Williams, Robinson\\
& Gilgeous-Alexander kept & $-0.08$ & Leonard, Bogdanovi\'c, Grant, Noah & Robinson\\
\midrule
Brooklyn, 2020 & engine's plan (126.74) & --- & Harden, Green, Shamet & Dinwiddie, LeVert, Prince, Kurucs\\
& Harden ruled out & $-5.75$ & Ingram, Green, Vonleh & Dinwiddie, LeVert, Prince, Jordan\\
& Allen sent away & $-0.93$ & Harden, Green, Vonleh & Allen, LeVert, Prince, Kurucs\\
& Dinwiddie kept & $-0.32$ & Harden, Green, Shamet, Vonleh & LeVert, Prince, Jordan\\
\bottomrule
\end{tabularx}
\end{table}

\subsection{Cleveland, 2017: the price of a trade request}

The 2017 Cleveland Cavaliers illustrate how the engine quantifies the hidden cost of an unexpected trade request. In July 2017, Kyrie Irving asked to be traded, and Cleveland sent him to Boston for Isaiah Thomas, Jae Crowder, and draft picks; it also signed Derrick Rose and, later in the offseason, Dwyane Wade. Free to keep Irving, the engine does so. It acquires Thomas and Crowder by trading away JR Smith's salary, signs Rose, and builds a roster worth about 1.3 points per 100 possessions more than Cleveland's. Boston, however, gave up Thomas and Crowder only to receive Irving, which the engine's market cannot know. Once Irving has to leave, as he did, the engine's best plan brings in the four players at the heart of Cleveland's summer: Thomas, Crowder, Rose, and Wade. In that constrained re-solve, the engine's plan matches the value of Cleveland's realized roster. Given the trade request, the model agrees with the front office's choices, and quantifies the offensive cost of accommodating the request at approximately 1.3 points per 100 possessions (Figure~\ref{fig:moves}). Of the players Cleveland received, Thomas was the one that mattered: without him the plan loses almost four points, while Crowder is worth about half a point.

\subsection{The Clippers, 2019: a shooter beside Leonard}

\begin{figure}[tbp]
\centering
\includegraphics[width=\textwidth]{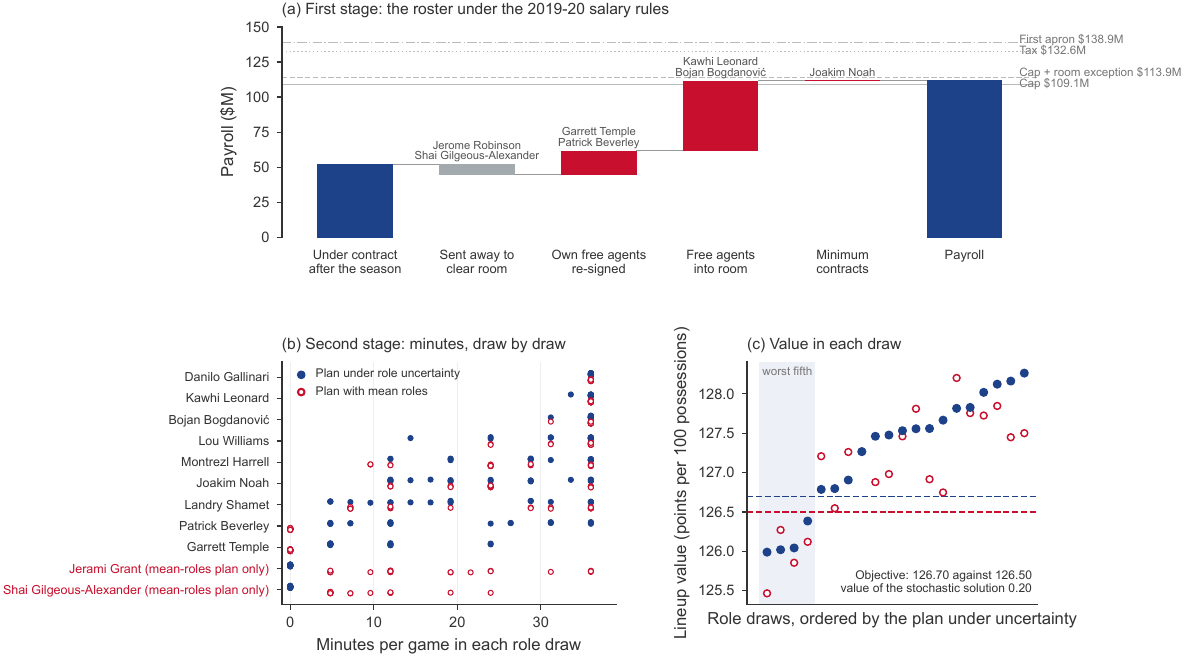}
\caption{The two-stage program on the Clippers' summer of 2019. (a) First stage: the payroll through the cap-room phase
under the 2019-20 rules. (b) Second stage: each player's minutes in each of the 20 role draws, for the plan made under
role uncertainty (filled) and the plan made with each player's average role (open). (c) Each plan's value in each draw;
the dashed lines are the two objectives.}\label{fig:case}
\end{figure}

The 2019 Los Angeles Clippers reveal how pairing two high-usage creators can congest offensive shot distribution. In July 2019, the Clippers signed Kawhi Leonard and traded Shai Gilgeous-Alexander, Danilo Gallinari, and five first-round draft picks to Oklahoma City for Paul George. Given identical market alternatives and salary-cap rules, the engine also signs Leonard, but it declines to trade for George (Figure~\ref{fig:case}a). It clears cap space by sending away Gilgeous-Alexander and Jerome Robinson, re-signs Patrick Beverley and Garrett Temple, signs Leonard and the shooter Bojan Bogdanovi\'c into that space, adds Joakim Noah at the minimum salary, and retains Gallinari and Lou Williams. The model values this roster about three and a half points per 100 possessions above the roster the Clippers fielded.

The engine rejects the George trade because adding another primary creator yields diminishing returns on a roster already rich in shot creation. Leonard and George are both primary creators, and the Clippers already had two more in Gallinari and Lou Williams. Because basketball possessions are finite, creators take shots from one another, so a second star creator adds much less to this roster than his own numbers suggest, while a shooter adds what the roster lacks. Marginal transaction values highlight this trade-off (Figure~\ref{fig:moves}). When Bogdanovi\'c is ruled out, the engine does pair Leonard with George, sending Lou Williams away to afford him, and the plan loses two points; George in place of Leonard loses almost one.

The second stage matters as much as the first. Figure~\ref{fig:case}(b) tracks each player's minutes across the twenty posterior role draws: Landry Shamet and Patrick Beverley play anywhere from five to more than thirty minutes a game, depending on how the roles turn out. A rotation fixed in advance for all draws, the single-stage program of Section~\ref{sec:model-program}, would lose 2.2 points, the largest value of flexible minutes in the three summers. The plan made for each player's average role keeps Gilgeous-Alexander for a few minutes and adds Jerami Grant in place of Beverley and Temple; it is worth a fifth of a point less and falls short in 15 of the 20 draws (panel c).

One move in the engine's plan is almost free to undo. Keeping Gilgeous-Alexander costs less than a tenth of a point, because the engine sends him away only to make cap space. A front office that valued his future, which a model of the coming season does not price, could have kept the player who went on to become the league's most valuable player at almost no cost to its offense.

\subsection{Brooklyn, 2020: why Harden}

The 2020 Brooklyn Nets demonstrate how micro-level shot charts distinguish stars who share identical categorical roles. Kevin Durant joined Brooklyn in 2019 but missed the 2019-20 season recovering from a torn Achilles tendon, so the engine values him on his 2018-19 season. In January 2021, the Nets traded Caris LeVert, Jarrett Allen, Taurean Prince, Rodions Kurucs, and draft picks for James Harden. Offered Harden as a trade target in the summer, the engine makes the trade as well, with one difference: it sends Spencer Dinwiddie instead of Allen and keeps Allen as its center. It also adds Jeff Green and Landry Shamet, and its rotation gives Kyrie Irving 24 minutes a game beside Harden and Durant.

\begin{figure}[tbp]
\centering
\includegraphics[width=\linewidth]{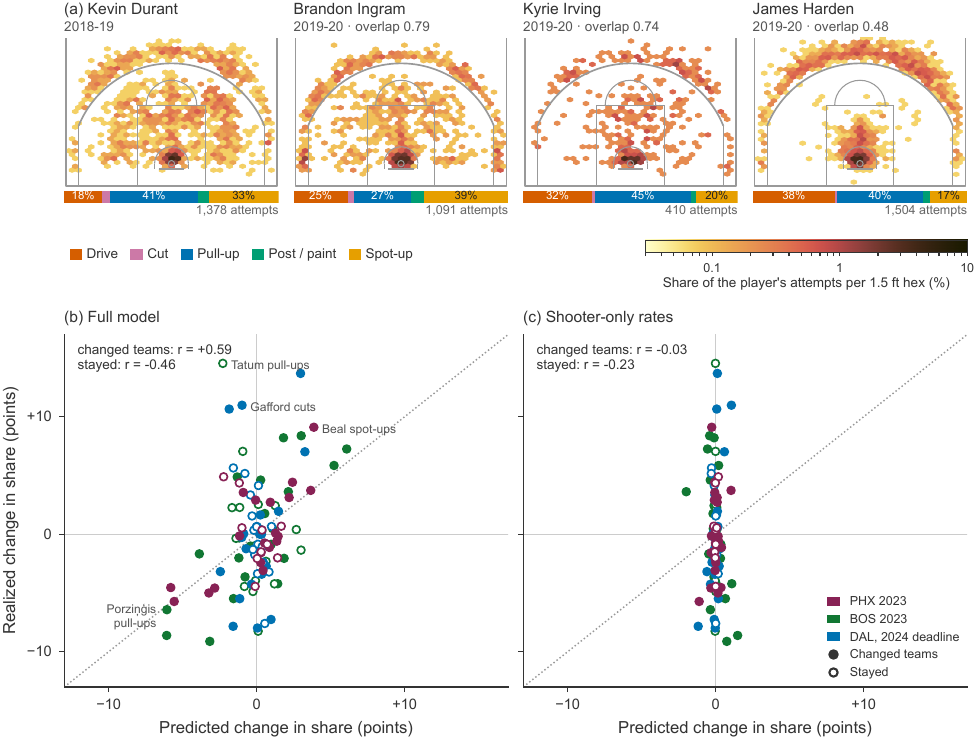}
\caption{The shot patterns behind the case studies. (a) Kevin Durant's shots in 2018-19 beside those of three creators
Brooklyn could pair with him, in 2019-20. Color: the share of the player's attempts in each hexagon; bars: his mix of
movement types. Overlap: the share of shots that the two players' 25-cell mixes have in common (1 for identical mixes).
(b, c) Predicted against realized changes in the shot mixes of eleven key players of three moves (Phoenix and Boston in
2023, Dallas at the 2024 trade deadline), by movement type and zone, under the full model (b) and shooter-only rates
(c), all scored with the model fitted on 2022-23.}\label{fig:caseshots}
\end{figure}

Spatial shot distribution explains why Harden complements Durant far better than other elite scoring creators. Durant, Harden, Irving, and Brandon Ingram, the best alternative in Brooklyn's market, are all primary creators in the model, so categorical roles alone cannot separate them. Their shot charts resolve the question (Figure~\ref{fig:caseshots}a). Ingram's and Irving's charts resemble Durant's, with many shots from mid-range; Harden's does not, because he scores at the rim and from beyond the arc. Measured by the share of shots that two charts have in common, Harden overlaps Durant by 0.48, against 0.74 for Irving and 0.79 for Ingram. Players shoot less where their teammates shoot, so Harden keeps more of his value beside Durant than the others would, and he had led the league in scoring as well. Without him, the engine turns to Ingram and its plan loses almost six points, by far the largest value of any single move in the three summers. Sending Allen away, as the Nets did, costs almost a point and leaves a roster worth about the same as the one Brooklyn fielded. In the engine's reading, the Nets chose the right star (James Harden) but paid for him with the wrong player (Jarrett Allen).

\subsection{What the model does not see: Boston, 2023}

The summer of 2023 shows both what a model of offense built from roles sees and what it does not. Milwaukee traded Jrue Holiday for Damian Lillard, a move the engine's plan for the Bucks makes as well, and the Bucks' offensive rating rose by 3.0 points, close to the calibrated forecast of 3.4 points for the lineups they used. Boston traded Marcus Smart for Kristaps Porzi\c{n}\c{g}is and sent Malcolm Brogdon and Robert Williams III to Portland for Holiday; the following season it had the league's best offense and won the title. Its offensive rating rose by more than five points, yet the model saw almost no gain in the lineups Boston used. Part of the difference lies outside the model by design, since Holiday was prized above all for his defense. The rest is instructive. The two newcomers took fewer pull-ups beside their new teammates, as predicted, but Jayson Tatum, who stayed, took many more of his shots off the dribble, where the model expected him to defer to the newcomers. Phoenix, which traded Chris Paul, Landry Shamet, and draft picks that summer for Bradley Beal to play beside Devin Booker and Kevin Durant, tells the same story. Beal and Durant, who had played most of the previous season elsewhere, gave up as many pull-ups as the model predicted and took more spot-ups, while Booker took more pull-ups than before. Across the eleven key players of three such moves, Phoenix and Boston in 2023 and Dallas at the 2024 trade deadline, the predicted shifts track the realized ones for the six players who changed teams ($r=0.59$) and run against them for the five who stayed ($r=-0.46$; Figure~\ref{fig:caseshots}b), while shooter-only rates predict almost no shift at all (panel c). The asymmetry points to offensive hierarchy as an additional determinant of how players adjust to a new roster, with newcomers adapting to their teammates while established stars retain offensive priority. Roles describe how players play, but not whose role adjusts when the roster changes, pointing to a natural extension of the model.
\section{Conclusions}\label{sec:conclusion}

Roster construction in professional basketball is a decision about
combinations. We have shown that the way teammates shape each other's
shots can be estimated from public micro-level shot data, turned into
the value of any lineup, and built into a risk-averse two-stage
stochastic program that respects the league's salary rules. Tested
against every team's offseason over eight seasons, and on two seasons
held out until the method was fixed, the engine finds rosters that are
clearly stronger than those the teams built.

The engine is best read as a decision support tool, and its results
carry five implications for how such a tool should be used in
practice. First, value players in context rather than by individual
box scores: a front office that ignores teammate effects would build a
different roster in almost every summer and give up more than a point
per 100 possessions, because these interactions are first-order
effects on a roster's offense. Second, plan for the average role but
keep the minutes flexible: in-season rotation changes absorb most of
the uncertainty about how newcomers will be used (for the 2019
Clippers, fixing the rotation in advance would have cost more than two
points per 100 possessions), and explicit hedging pays only when a
roster's value depends on several players whose roles are unsettled.
Third, treat the optimal roster as a short list: the objective
landscape is flat near the optimum, the best plan that kept
Gilgeous-Alexander cost less than a tenth of a point per 100
possessions, and the decision maker adds most value by choosing among
near-optimal rosters on dimensions the model does not observe. Fourth,
price the moves, not only the roster: re-solving the program with a
single move imposed or ruled out yields the decision value of each
option, turning sensitivity analysis into a negotiation tool.
Fifth, protection against playoff defenses is cheap: a roster built to withstand a strong playoff defense that cuts into two of its favorite shot types gives up only a few hundredths of a point per 100 possessions in the regular season, and the stress test can quantify each candidate roster's exposure before the front office commits to its acquisitions.

Several limitations bound the scope of these implications. The engine
values offense only; the trade market is modeled generously; the
reported gains are measured by the same model that found them and are
therefore subject to the optimizer's curse \citep{smith2006},
mitigated but not eliminated by out-of-sample validation; the stress
test measures exposure within the model rather than predicting actual
playoff declines; and roles are forecast from the season just ended,
so the model foresees neither injuries nor the value of draft picks.
Each limitation suggests a direct extension: a defensive value
function constructed in the same way from the shots a lineup's
opponents take; explicit modeling of the other side of each trade;
linking stress-test exposure to realized playoff outcomes over a
longer panel; and embedding the single-offseason model in a
multi-season dynamic program with injuries, aging, draft assets, and
contract years as state variables. The same framework also extends naturally to workforce planning and other team-assembly problems in which individual productivity depends on colleagues and managers can adapt operational assignments after personnel commitments are made.

\bibliographystyle{plainnat}
\bibliography{references}

\newpage
\appendix
\let\appendixsection\section
\renewcommand{\section}{\FloatBarrier\appendixsection}   
\begin{center}
{\Large\bfseries Appendix}
\end{center}
\section{Data}\label{app:data}

\begin{figure}[ht]
\centering
\includegraphics[width=0.85\textwidth]{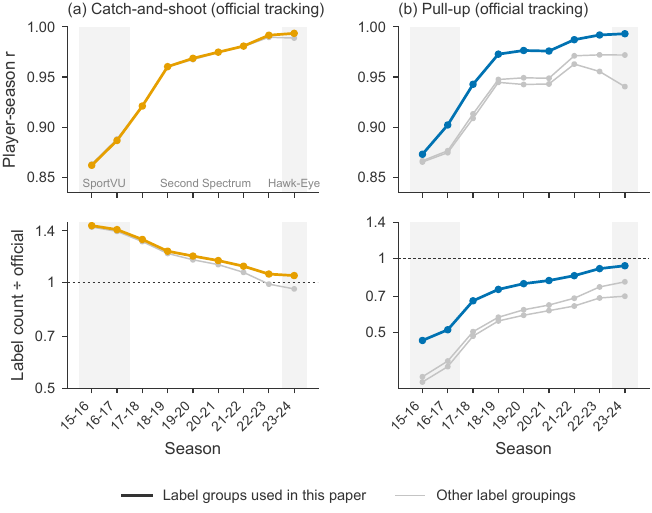}
\caption{Movement labels against the league's tracking classifications, by season.}\label{fig:labels}
\end{figure}

\begin{figure}[ht]
\centering
\includegraphics[width=0.85\textwidth]{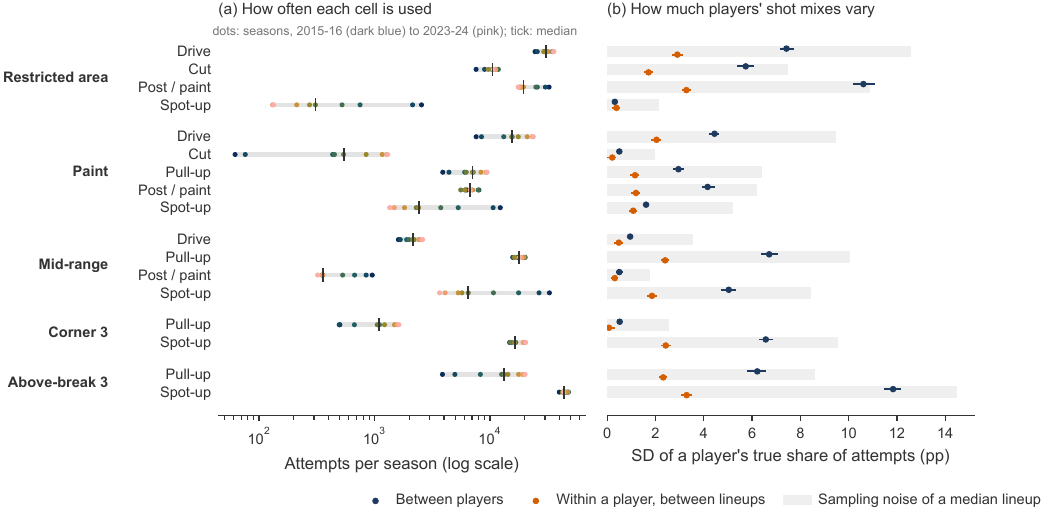}
\caption{Attempts per cell, and the variation of shot mixes between and within players.}\label{fig:support}
\end{figure}

\paragraph{Shots and cells.} Every regular-season field-goal attempt carries the scorer's coordinates and an action
description. The zone follows from the coordinates (restricted area, paint outside it, mid-range, corner three,
above-the-break three); the movement type from the description, mapped to drive, cut, pull-up, post-up or paint touch,
and spot-up. Labels were checked against the league's tracking-based play-type classifications~\citep{nbastats} season by season
(Figure~\ref{fig:labels}); scorers in different arenas use the default labels at different rates, which is why the
shot model carries an arena-by-movement term. The same retention rule applies to every season.
Figure~\ref{fig:support} shows how often each cell is used and how much players' mixes vary within and between
players.

\paragraph{Lineups and exposures.} The unit of the shot model is a stint: an interval with the same ten players on the
floor. Lineups come from the league's rotation records or, where those are missing, from a reconstruction of the
substitutions in the play-by-play. Exposure is counted in offensive possessions.

\paragraph{Salaries and the cap.} Player salaries by season come from Basketball-Reference team payrolls~\citep{bbref}, and for 2024-25 and 2025-26 from its contracts page as archived after opening night~\citep{bbrefcontracts}, matched to
league identifiers by name, team and season; a player who logged minutes without a salary row receives the season's
minimum salary for zero years of service. A mover's salary skips the row of the team he left when it records a buyout.
Table~\ref{tab:cap} gives the amounts of each league year, each checked against a league release~\citep{nbapr} or contemporary
press source~\citep{hoopsrumors,bbinsiders}.

\begin{table}[ht]
\centering\footnotesize
\caption{Salary-cap amounts by league year (\$M), and the collective bargaining agreement in force~\citep{nbacba2011,nbacba2017,nbacba2023}.}\label{tab:cap}
\begin{tabular}{lccccccc}
\toprule
League year & Cap & Tax & First apron & Second apron & Mid-level & Taxpayer mid-level & Agreement\\
\midrule
2016-17 & 94.14 & 113.29 & 117.29 & --- & 5.63 & 3.48 & 2011\\
2017-18 & 99.09 & 119.27 & 125.27 & --- & 8.41 & 5.19 & 2017\\
2018-19 & 101.87 & 123.73 & 129.82 & --- & 8.64 & 5.34 & 2017\\
2019-20 & 109.14 & 132.63 & 138.93 & --- & 9.26 & 5.72 & 2017\\
2020-21 & 109.14 & 132.63 & 138.93 & --- & 9.26 & 5.72 & 2017\\
2021-22 & 112.41 & 136.61 & 143.00 & --- & 9.54 & 5.89 & 2017\\
2022-23 & 123.66 & 150.27 & 156.98 & --- & 10.49 & 6.48 & 2017\\
2023-24 & 136.02 & 165.29 & 172.35 & 182.79 & 12.41 & 5.00 & 2023\\
2024-25 & 140.59 & 170.81 & 178.13 & 188.93 & 12.82 & 5.17 & 2023\\
2025-26 & 154.65 & 187.90 & 195.95 & 207.82 & 14.10 & 5.69 & 2023\\
\bottomrule
\end{tabular}
\end{table}

\paragraph{Movers and transactions.} For each player who changed teams after season $t$, transaction records~\citep{bbref} give how
he left (traded in the summer, traded or waived during $t$, free agent) and his contract status in $t+1$. A summer
trade leaves the player under his contract; a team cannot re-sign a player it waived; a player gone during $t$ is no
longer his old team's.

\paragraph{Seasons.} Only the first pair reads 2015-16's play-by-play, shots and playoffs, and its role model adds the box-score totals of 2014-15; every other pair's role model reads 2015-16 through its box score only. In 2024-25 one game with a defective feed is excluded. In 2025-26 the league's traditional box-score service returns empty tables for most games; those games take the next
version of the service, written in the same format, after checking that the two agree on the games both return.

\section{The role model}\label{app:roles}

\paragraph{Model.} To value potential additions by player type, we define player roles through a Bayesian mixture model that joins box scores with shot charts. For player-season $i$ in season $s(i)$ with minutes $M_i$ and role $z_i$,
\begin{align*}
X_{ij}\mid z_i=k &\sim \mathrm{NegBin}\big((M_i/36)\,\xi_{s(i)j}\,\mu_{kj},\;\phi_j\big), \qquad j\in\{\text{box counts, total attempts}\},\\
Y_i\mid n_i, z_i=k &\sim \mathrm{DirMult}\big(n_i,\;\varkappa\,\mathrm{softmax}(\log\psi_k+\delta_{s(i)})\big),\\
z_i &\sim \mathrm{Categorical}(\pi_{\mathrm{group}(i)}),
\end{align*}
where $\xi$ is a season-by-statistic multiplier for league-wide drift, $Y_i$ is the count of the player-season's attempts over the
25 cells, $\psi_k$ represents role $k$'s cell shares, $\delta_s$ is a season shift per cell, and $\pi$ is the role mix of each position
group. The box counts are points, attempts, free throws, assists, turnovers, rebounds, steals and blocks. The
Dirichlet-multinomial term enters the likelihood with weight 0.1, so that shots refine roles defined by the whole box
score instead of overriding them. We sum out the discrete role assignments analytically and sample the remaining parameters with the No-U-Turn sampler
\citep{hoffman2014} in NumPyro \citep{phan2019} on a GPU.

\begin{figure}[ht]
\centering
\includegraphics[width=0.85\textwidth]{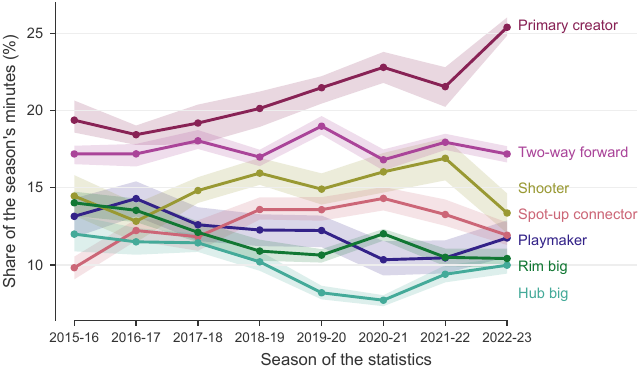}
\caption{The seven roles' share of minutes by season, with 90\% posterior bands.}\label{fig:rolemix}
\end{figure}

\paragraph{Choosing the number of roles.} We choose seven roles for their use to a front office rather than for forecasting accuracy. To select the number of roles, we fit a sparse mixture of ten components to every season
under the coarsened posterior of \citet{miller2019}, with robustness parameter $\alpha\in\{50,100,150,200,300,500\}$
across four chains each. This sweep yields 4, 5, 6, 6, 9 and 10 occupied roles. The fit gained per step falls from 1.47 to 0.33 by
six roles, the smallest role holds 14.0\% to 6.2\% of minutes, and chain agreement is highest up to six or seven roles. Because no
value in the grid yields seven or eight roles, we fit seven roles directly. For each of the seven pairs whose training season falls between 2016-17 and 2022-23, all four chains of the exact posterior find seven roles, with role-rate $\hat R$ at most 1.013. The seventh role splits the forwards into spot-up connectors (corner
spot-ups and rim cuts) and two-way wings, a distinction front offices use. At full shot weight, the runner-up configuration with six roles costs 0.10 to 0.13 per game of held-out log score. Figure~\ref{fig:rolemix} shows each role's share of minutes by season.

\paragraph{Fold fits and draws.} We fit the roles separately for each season pair and draw from the posterior to represent uncertainty in how players will be used. Each season pair's roles are fitted on the seasons up to its training season. The engine's roles and draws come from the coarsened posterior ($\alpha=200$), whose spread captures a player's role uncertainty; Figure~\ref{fig:rolemix} uses the exact posterior of the pair trained on 2022-23. Under the coarsened posterior the seventh role occasionally empties in a draw,
which loosens the role rates of the kept chains. Players with fewer than 100 minutes in the season before a decision
draw their roles from their position group's role mix with a prior weight of three, so their uncertainty is wide. The
same 20 draws are used by every program of a season. Across the ten fits, one to three of the four chains share the reference chain's roles (the others split the same continuum of players differently), with no divergences and role-rate $\hat R$ between 1.03 and 3.21, the highest for the fit through 2018-19. The first pair's fit, on 2015-16 and the box-score totals of 2014-15, keeps three chains, with 8 of 50 draws leaving a role empty ($\hat R$ 1.11), and the fit through 2024-25 keeps one, with 13 of 50 ($\hat R$ 1.41). All fits were used as fitted.

\section{The shot model}\label{app:shot}

\paragraph{Attempts.} To quantify how teammates influence where and how frequently a player shoots, we model attempt counts across court cells with a penalized Poisson regression. For the player in lineup slot $u$ of a stint, the attempt count $Y_{uc}$ in cell $c$ is Poisson
with mean $E_u\exp(\eta_{uc})$, where $E_u$ is the stint's offensive possessions and
\[
\eta_{uc}=\mu_c+\tfrac12 h\,\mathds{1}[\text{home}]+\theta_{\mathrm{arena}(u),m(c)}+s_{q(u),c}+\beta_c^{\top}b_u
+(a_u^{\top}u_{g(c)})(b_u^{\top}v_{g(c)}),
\]
with $b_u=H^{\top}\bar w_{-u}$ the four teammates' mean role shares and $a_u=H^{\top}w_u$ the shooter's own, both in
Helmert contrasts $H$ of the seven roles. The arena term absorbs how each arena's scorers record movement types. Because
the role shares sum to one, only contrasts of the slopes are identified, and we report them as relative shifts against
an average teammate \citep{li2023}. We fit the model as a ridge-penalized Poisson regression at its
posterior mode under normal priors: players' profiles are pulled toward the league average so that a player with few
shots is not given an extreme profile, and the shooter profiles $s$ and arena terms $\theta$ have their own ridge penalties. The precisions of the teammate effects (the role slopes, their pull toward the cell cluster and the interaction terms) are chosen by the marginal likelihood on season $t$; the profile penalty is the one that forecasts best on the previous season pair (fitted on $t-1$ and scored on $t$; for the first pair, on a split of its training season's games); and the arena penalty is fixed at 4. No data from season $t+1$ enter any penalty.

\paragraph{Makes.} To determine whether an attempt succeeds, we fit a logistic regression model with the same teammate predictors:
\[
\mathrm{logit}\,p_{qc}=\mu^{m}_c+s^{m}_{qc}+\beta^{m\top}_c b_q .
\]

\begin{figure}[ht]
\centering
\includegraphics[width=0.95\textwidth]{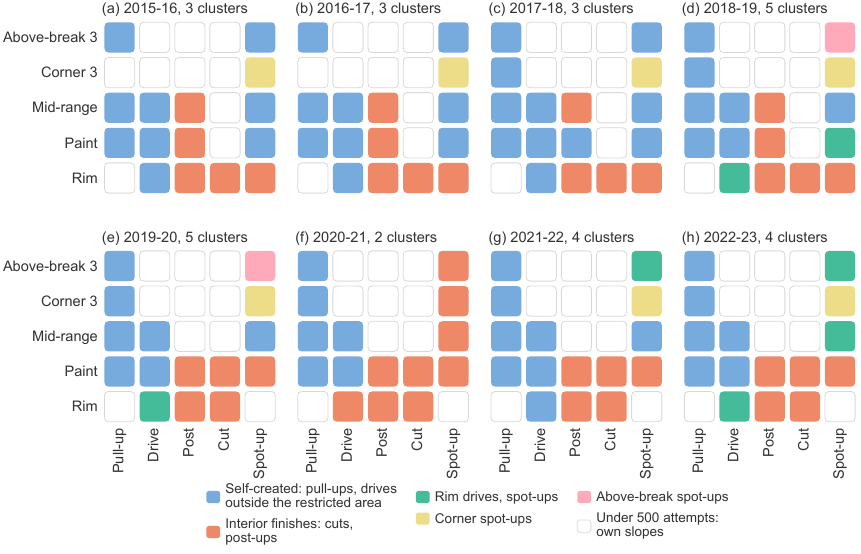}
\caption{Groups of cells that share their teammate effects, one panel per season. Each square is a zone (rows) and a
movement type (columns); cells with the same color share their effects, and white cells, with fewer than 500 attempts,
keep their own.}\label{fig:clusters}
\end{figure}

\paragraph{Sharing across cells.} We share statistical strength across court locations by clustering cells that exhibit similar responses to teammates. At the fitted coefficients, the curvature of the log likelihood in the role block,
with the other blocks profiled out, gives a normal approximation $b_c\sim N(\beta_c,\Lambda_c^{-1})$ of each cell's
six teammate-role slopes. The cells with at least 500 training attempts are clustered by the MRF-constrained mixture
of finite mixtures \citep{miller2018} of \citet{meng2025}, run on these approximations (neighbors within graph distance
two on the zone-by-movement grid; $\lambda$ chosen by the log pseudo-marginal likelihood; Dahl's estimate of the
partition, \citealp{dahl2006}); the clustering favors grouping cells that are neighbors on the court. Each cell's slopes
are then pulled toward its cluster mean, $\beta_c\sim N(\bar\beta_{g(c)},\tau^{-1}I)$, with $\tau$ chosen by the
marginal likelihood of the approximations; the nine or ten cells with fewer attempts keep their own slopes. The partitions recur
across seasons: self-created shots (drives and pull-ups), interior finishes (cuts and post-ups) and corner spot-ups form
clusters in most pairs (Figure~\ref{fig:clusters}). Against independent cells whose ridge is chosen the same way, the
pulled slopes gain $+0.00$ per game ($+0.00$ to $+0.01$) and are better in five of the eight pairs; fixing the
independent cells' ridge at 4 instead raises their gain from $+2.05$ to $+2.07$.

\paragraph{The interaction ladder.} We test higher-order interactions to evaluate whether a player's own role changes how he responds to teammates. Over the additive model, the rank-one shooter-by-teammate interaction gains $+0.07$
per game on average (95\% interval $+0.03$ to $+0.11$; $-0.09$ to $+0.26$ across pairs), a rank-two interaction and teammate-pair redundancy terms nothing more, and the full-rank
interaction $+0.09$; five pairs gain and three lose a little. The makes model carries teammates' roles as well; they
add nothing to the held-out score, so the teammate effects work through where and how often players shoot.

\paragraph{The overlap term.} We incorporate a crowding term to account for direct competition when teammates seek shots in the same court cell. With the interaction model as an offset, the term $\gamma\,O_{qc}$, where $O_{qc}$ is the
four teammates' expected attempts in the player's cell per 100 possessions, is fitted on season $t$ with cell
intercepts and scored on $t+1$. Its coefficient is negative in every pair ($-0.0080$ to $-0.0188$ per teammate attempt,
standard errors about 0.001) and its held-out gain positive in all eight ($+0.13$ to $+0.35$ per game); a term in the
teammates' total usage adds nothing once it is in.

\paragraph{Transportability.} We verify that predicted teammate effects carry over to players who change teams. Figure~\ref{fig:forecast}(a, b) gives the damping factor $\kappa$ across the court and by zone. On the movers'
held-out shot mix, the additive model gains $+0.0045$ per attempt over shooter-only rates ($+0.0028$ to $+0.0060$) and
the mover's own season-$t$ mix loses $0.022$.

\section{Salary-rule constants}\label{app:rules}

We state the trade-matching constants and salary aggregation rules of the 2011, 2017 and 2023 collective bargaining agreements~\citep{nbacba2011,nbacba2017,nbacba2023} that complete the cap constraints of Section~\ref{sec:model-program}. The salary-cap constraints \eqref{eq:roomlink}--\eqref{eq:trades} use the amounts of each league year in
Table~\ref{tab:cap} and the trade-matching constants below. The narrow rule $I\le a_0O+b_0$, which applies above the
line, has $(a_0,b_0)=(1.25,\$0.1M)$ under the 2011 and 2017 agreements, $(1.10,0)$ in 2023-24 and $(1.00,0)$ from
2024-25. Under the line, the ceiling $a_1O+b_1$ and the two bands $O+b_2$ and $a_3O+b_3$ have
$(a_1,b_1,b_2,a_3,b_3)=(1.75,0.1,5,1.25,0.1)$ under the 2017 agreement, $(1.5,0.1,5,1.25,0.1)$ under the 2011 agreement and
$(2.0,0.25,7.5,1.25,0.25)$ under the 2023 agreement, in \$M. Starting in 2024-25, a team that aggregates two or more outgoing salaries in a
trade is capped at the second apron. One more binary variable $z^{\mathrm{agg}}$ records this, with
$S_0+\sum_{i}S_ik_i\le\mathrm{Apron}_2+M(1-z^{\mathrm{agg}})$ and, for every trade arrival $j$,
\[
z^{\mathrm{agg}}\ge k_j-x_j-\frac{1}{|\Rt|-1}\sum_{i\in\Rt}(k_i+x_i),
\]
which forces $z^{\mathrm{agg}}=1$ when a player arrives in the trade phase while two or more players under contract
leave in it.

\section{Stress-test calibration}\label{app:stress}

\begin{figure}[h]
\centering
\includegraphics[width=0.85\textwidth]{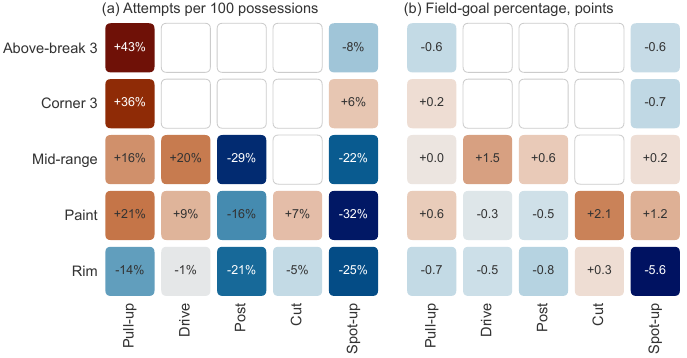}
\caption{The playoff shift: playoff attempts per 100 possessions against the league's regular-season rates, by zone and movement type,
2016-17 to 2023-24.}\label{fig:playoffshift}
\end{figure}

Strong defenses are each season's playoff teams in the best quarter by regular-season field-goal points allowed per
100 possessions. For each series against them, the offense's expected attempts per cell are its own regular-season
rates times the series' possessions; the cut in a cell is one minus playoff over expected attempts, kept between 0 and
0.5. The removed attempts that reappear in other cells, and the points they score there, give the value $v^{\mathrm{rr}}$
of a removed attempt. Each decision season uses the playoffs up to its training season only (Table~\ref{tab:stress}).
Over the 62 series of 2016-17 to 2022-23, strong defenses cut rim cuts by 17\%, rim post-ups by 21\%, paint spot-ups by
23\%, mid-range spot-ups by 20\% and above-the-break spot-ups by 10\%; above-the-break pull-ups rise 34\%, paint drives
25\% and mid-range pull-ups 10\%. For the broader playoff shift, Figure~\ref{fig:playoffshift} compares all playoff games with the league's regular-season rates over 2016-17 to 2023-24.

\begin{table}[h]
\centering\small
\caption{Calibration by training season: series against strong defenses, mean cut, share of removed attempts re-routed,
and value of a removed attempt.}\label{tab:stress}
\begin{tabular}{lcccc}
\toprule
Training season & Series & Mean cut & Re-routed & Value of a removed attempt\\
\midrule
2015-16 & 6 & 0.20 & 1.00 & 1.01\\
2016-17 & 10 & 0.08 & 0.96 & 1.23\\
2017-18 & 19 & 0.14 & 0.93 & 1.03\\
2018-19 & 27 & 0.13 & 0.88 & 0.92\\
2019-20 & 38 & 0.12 & 0.82 & 0.80\\
2020-21 & 44 & 0.12 & 0.85 & 0.82\\
2021-22 & 57 & 0.09 & 0.83 & 0.78\\
2022-23 & 62 & 0.09 & 0.87 & 0.81\\
2023-24 & 69 & 0.09 & 0.85 & 0.79\\
\bottomrule
\end{tabular}
\end{table}

\section{The frozen protocol and its deviations}\label{app:protocol}

\paragraph{What was frozen.} Before either untouched season was opened, the code was fixed at one commit; the models
(seven shot-informed roles, the 25-cell shot model with pulled cell clusters, the rank-one interaction and the overlap
term, the capped lineup value with $\kappa=0.75$, the engine's objective and constraints, the stress calibration); the
data rules; and the ten claims of Table~\ref{tab:sealed}, each with its range over the eight open pairs. Each untouched
season was then processed and scored once by a scripted run in which every fit, penalty and tuning choice uses the
seasons before it, and the season enters only to be scored. The run was rehearsed beforehand on the 2022-23 $\to$
2023-24 pair under another name; it reproduced that pair's quantities to within 0.01, except the value of modeling
synergy (0.48 against 0.44, and 1.07 against 1.16), because the engine chose differently among rosters of equal value
(the same roster for 17 of 30 teams, values within 0.007 for all).

\paragraph{Three deviations.} Three data problems came to light during the runs. Each fix changes how data are read, not
a model, a setting or a claim, and each was tested before its results were used.
\begin{enumerate}[label=(\roman*)]
\item \emph{Minimum salaries for the untouched seasons.} The salary table's rule for players without a salary row had
not run for 2024-25, whose box scores were sealed when the table was built, so 27 of the 139 movers into 2024-25 had no
salary and 24 of the 30 pools could not be built. The first run solved six offseasons (engine minus real roster 4.90,
VSS 0.05, synergy 0.31 and 0.77, value of overlap 0.12 on those six). With the rule applied (156 rows for 137 players),
the re-run solved 29.
\item \emph{2025-26 box scores.} The league's traditional box-score service returned no rows for 1,199 of the 1,230 games;
those games take the next version of the service, in the same format. On the 31 games both return, the two agree
except for one plus-minus value among 799 player rows.
\item \emph{The order of games.} A player's last team in season $t$ and first team in $t+1$ were read from box scores
sorted by game identifier, which is not chronological where games were rescheduled (2020-21, 2021-22) or added late
(the in-season tournament, from 2023-24); some players acquired at the trade deadline fell out of their new team's
pool. Ordering by game date changes 13 of the 30 pools of 2025, 8 of 2024 (including Washington, which without its
deadline acquisitions had no feasible roster) and 12 of the open pairs (5 in 2020, 4 in 2021, 3 in 2023). Every
affected case was solved again in every version it had been run in. Table~\ref{tab:deviation} gives claims 7 to 10
before and after; of the open-pair ranges, two move at two decimals: claim 7's lower end (3.68 to 3.67) and claim 8's upper end (0.46 to 0.45).
\end{enumerate}

\begin{table}[h]
\centering\small
\caption{Claims 7--10 before and after the game-order fix, for the offseasons whose pools changed.}\label{tab:deviation}
\begin{tabular}{lcccc}
\toprule
Offseason & Engine minus real & VSS & Synergy (additive / shooter-only) & Overlap\\
\midrule
2020 & 4.05 $\to$ 4.04 & 0.29 $\to$ 0.27 & 0.58 / 1.82 $\to$ 0.53 / 1.84 & 0.03 $\to$ 0.02\\
2021 & 4.65 $\to$ 4.65 & 0.46 $\to$ 0.45 & 0.46 / 0.96 $\to$ 0.48 / 0.96 & 0.09 $\to$ 0.09\\
2023 & 3.68 $\to$ 3.67 & 0.17 $\to$ 0.17 & 0.44 / 1.16 $\to$ 0.43 / 1.11 & 0.06 $\to$ 0.05\\
2024 (2024-25) & 4.49 $\to$ 4.50 & 0.05 $\to$ 0.07 & 0.47 / 0.93 $\to$ 0.58 / 0.91 & 0.09 $\to$ 0.10\\
2025 (2025-26) & 3.94 $\to$ 3.83 & 0.18 $\to$ 0.14 & 0.78 / 1.48 $\to$ 0.82 / 1.49 & 0.09 $\to$ 0.07\\
\bottomrule
\end{tabular}
\end{table}

\paragraph{Role fits used as fitted.} The role fit through 2024-25, which feeds the 2025-26 scores, kept one of its four
chains (the others found six roles); 13 of its 50 draws leave a role empty and the role rates' $\hat R$ reaches 1.41.
It was used as fitted, as were the other fits (Appendix~\ref{app:roles}).

\end{document}